\documentclass[trackchanges,twocolumn,twocolappendix]{aastex701}
\usepackage{multirow}

\begin{document}

\title{SMA 200–400 GHz Survey for 10 faint Class II Disks in the Taurus Molecular Cloud}

\author[0009-0007-3677-8040]{Chia-Ying Chung}
\affiliation{Department of Physics, National Sun Yat-Sen University, No. 70, Lien-Hai Road, Kaohsiung City 80424, Taiwan, R.O.C.}
\affiliation{Department of Physics and Astronomy, Johns Hopkins University, 3400 N Charles Street, Baltimore, MD 21218, USA}  
\email{sharon0311chung@gmail.com}

\author[orcid=0000-0003-2300-2626]{Hauyu Baobab Liu}
\affiliation{Department of Physics, National Sun Yat-Sen University, No. 70, Lien-Hai Road, Kaohsiung City 80424, Taiwan, R.O.C.}
\affiliation{Center of Astronomy and Gravitation, National Taiwan Normal University, Taipei 116, Taiwan}  
\email[show]{hyliu.nsysu@g-mail.nsysu.edu.tw}
\correspondingauthor{Hauyu Baobab Liu}

\author[0000-0003-2253-2270]{Sean M. Andrews}
\affiliation{Center for Astrophysics \textbar\, Harvard \& Smithsonian, 60 Garden St., Cambridge, MA 02138, USA}
\email{sandrews@cfa.harvard.edu}

\author[0000-0003-0685-3621]{Mark A. Gurwell}
\affiliation{Center for Astrophysics \textbar\, Harvard \& Smithsonian, 60 Garden St., Cambridge, MA 02138, USA}
\email{mgurwell@cfa.harvard.edu}

\author[0000-0002-9154-2440]{Melvyn Wright}
\affiliation{Department of Astronomy, University of California, Berkeley, 501 Campbell Hall, Berkeley, CA 94720-3441, USA}
\email{melvyn.wright@gmail.com}

\author[0000-0002-7607-719X]{Feng Long}
\affiliation{Kavli Institute for Astronomy and Astrophysics, Peking University, 5 Yiheyuan Road, Haidian District, Beijing, 100871, People's Republic of China}
\email{long.feng@pku.edu.cn}

\begin{abstract}
We have performed Submillimeter Array (SMA) observations of 198.0--407.5 GHz broadband spectra of 10 Class II protoplanetary disks in the Taurus-Auriga region.
The 337 GHz flux densities ($F_{\rm 337 GHz}$) of these objects are in the range of 8.2--34 mJy.
The median and standard deviation of the 198--358 GHz spectral indices ($\alpha_{198-358}$) of these 10 Class II disks are 1.9 and 0.3, respectively.
Compared to the recent, similar SMA survey on another 47 Class II disks that are brighter at (sub)millimeter bands ($F_{\rm 337 GHz}\sim$20--730 mJy), there is no evidence that these newly observed 10 fainter Class II disks have systematically different $\alpha_{198-358}$ values.
At $>$230 GHz frequencies, the optical depths of these 10 fainter Class II disks may be as high as those of the brighter sources, which may be $\gtrsim$5.
In addition, the low ($<$2.0) values of $\alpha_{198-358}$ in some faint objects may be explained by the effect of dust self-scattering, with maximum dust grain size ($a_{\rm max}$) $\sim$ 100 $\mu$m, or by contamination by free-free emission.
\end{abstract}

\keywords{\uat{Circumstellar dust}{236} --- \uat{Protoplanetary disks}{1300} --- \uat{Pre-main sequence}{1289} --- \uat{Planet formation}{1241}}


\section{Introduction} \label{sec:intro}
The dust mass budget and dust grain sizes in protoplanetary disks (PPDs) crucially influence planet formation scenarios (for a review see \citealt{Birnstiel2024ARA&A..62..157B}).
(Sub)millimeter observations are efficient tools for constraining these quantities (for a review see \citealt{Andrews2020ARA&A..58..483A}).

\begin{deluxetable*}{lccccc}
\tabletypesize{\footnotesize}
\tablecolumns{8}
\tablewidth{0pt}
\tablecaption{Properties of the host protostars and the (sub)millimeter disk \label{tab:sources}}
\tablehead{ 
\colhead{Source} & \colhead{2MASS name} & \colhead{Distance} & \colhead{Spectral Type} & \colhead{Multiplicity} & \colhead{Separation}   \\
                 &                      &  (pc)              &                         &                        &   ($''$)  
}
\startdata 
CIDA-1 & J04141760+2806096 & 134.6 & M4.5 & single & $\cdots$ \\
FZ~Tau & J04323176+2420029 & 129.2 & M0.5 & binary & $\sim16$     \\
V410~X-ray~2 & J04183444+2830302 & 128.9 & M0 & unknown & $\cdots$  \\
CX~Tau & J04144786+2648110 & 126.7 & M2.5 & single & $\cdots$ \\
V807~Tau & J04330664+2409549 & 184.1 & K7 & triple & $\lesssim$0.3 \\
FX~Tau & J04302961+2426450 & 157.0 & M1 & binary & 0.89 \\
IT~Tau~A/B & J04335470+2613275 & 160.4 & K6/M2.9 & binary & 2.41 \\
KPNO~10 & J04174955+2813318 & 135.8 & M5 & single & $\cdots$ \\
04301+2608 & J04331435+2614235 & 172.8 & M0 & unknown & $\cdots$ \\
\enddata
\tablecomments{Except for V410~X-ray~2, the distances were quoted from {\it Gaia} DR3 (\citealt{Gaia_2016,Gaia_2023}). We quote the distance of V410~X-ray~2 from \citet{Akeson_2019}, which is the mean {\it Gaia} distance of the sources that are separated from it by less than 30$'$.
}
\tablerefs{Separation of binary and multiple systems were quoted from \citet{Andrews2013ApJ...771..129A,Akeson_2019}. Other source properties were quoted from \citet{Duchene1999,Luman2003ApJ...590..348L,Furlan2011ApJS..195....3F,Schaefer2012ApJ...756..120S,Herczeg2014ApJ...786...97H}}
\end{deluxetable*}

Previously, \citet{Chung2024ApJS..273...29C} observed the 200--400 GHz flux densities towards 47 Class II objects in the Taurus-Auriga region using the Submillimeter Array (SMA).
To achieve good signal-to-noise ratios with a limited amount of observing time, this previous SMA survey targeted sources that the flux densities at 337 GHz ($F_{\rm 337 GHz}$) are brighter than 20 mJy, which corresponds to the bright 32 percentile. 
They found that the 200--400 GHz spectral indices ($\alpha_{\rm 200-400}$) of most of the selected sources are populated in an extremely narrow range of $2.0 \pm 0.2$. 
The most straightforward interpretation of this result is that in the selected Class II samples, the (sub)millimeter luminosity is dominated by the {\it disk cores} that have high (sub)millimeter optical depths ($\tau\gtrsim$5), which is consistent with the previously reported size-luminosity relation (\citealt{Andrews2018ApJ...869L..41A,Tripathi2017ApJ...845...44T,Hendler2020ApJ...895..126H}).
Only some exceptionally spatially extended disks possess optically thin halos and thus have higher (sub)millimeter spectral indices.
Using population synthesis, \citet{Chung2024ApJS..273...29C} found that if assuming the standard DSHARP dust opacity (\citealt{Birnstiel2018ApJ...869L..45B}), the dust masses in their observed Class II disks may be, in general, greater than 100 $M_{\oplus}$ (see their discussion in Section 5.3.1).
This may be close to the initial dust mass budget of a passive protoplanetary disk (see below).
For example, theoretical calculations have shown that, for an isothermal, passive protoplanetary disk to be gravitationally stable, the gas and dust masses in the disk need to be less than 0.1 times the mass of the host protostar (\citealt{Kratter2016}).
Therefore, for a $\sim$0.3 $M_{\odot}$ protostar, the initial mass of a stabilized disk may be as high as 0.03 $M_{\odot}$.
Assuming that the gas-to-dust mass ratio is $\sim$100, then the initial dust mass budget would be 3$\times$10$^{-4}$ $M_{\odot}$, which is $\sim$90 $M_{\oplus}$.
A comparison of this estimate with the lower limits of dust masses constrained by \citet{Chung2024ApJS..273...29C} shows that the majority of dust mass may be retained during the Class II stage instead of being lost to forming planetesimals, (proto)planets, and/or protostars.

Moreover, \citet{Chung2024ApJS..273...29C} reported that in a few sources that have low $F_{\rm 337 GHz}$, the values of $\alpha_{\rm 200-400}$ are below 2.0.
In the Rayleigh-Jeans limit, when dust scattering opacity is negligible (e.g., when dust grain size remains small), in an optically thick dust structure, the lowest value of $\alpha_{\rm 200-400}$ that can be achieved by dust thermal emission is 2.0, which is consistent with that of the Planck function.
A lower than 2.0 value of $\alpha_{\rm 200-400}$ can be explained when the maximum dust grain size is close to $\lambda/2\pi$, where $\lambda$ is the observational wavelength. 
In this case, the attenuation of the observed flux density as a result of dust scattering opacity is not negligible.
In addition, when the maximum dust grain size is close to $\lambda/2\pi$, the dust albedo, which is the ratio between the scattering opacity and the total opacity, increases with observing frequency. 
This leads to more attenuation of the dust flux density at higher frequencies, which lowers the observed spectral indices, making it possible to be smaller than 2.0 (\citealt{Liu2019ApJ...877L..22L}).
For the observational wavelength range of \citet{Chung2024ApJS..273...29C}, this explanation of $\alpha_{\rm 200-400}<2.0$ corresponds to the maximum dust grain size of 120--240 $\mu$m.
Intriguingly, the recent ALMA dust polarization study on the PDS~70 disk has shown that, even after the presence of planets, the dust traced by (sub)millimeter observations may have $a_{\rm max}\lesssim$100 $\mu$m (\citealt{Liu2026arXiv260205247L}).

\begin{deluxetable*}{cccccccccc}
\tabletypesize{\scriptsize}
\tablecolumns{9}
\setlength{\tabcolsep}{1pt}
\tablewidth{0pt}
\tablecaption{SMA observations \label{tab:obs_summary}}
\tablehead{ 
\colhead{Track ID} & \colhead{UTC Date} & \colhead{LSB Freq.} & \colhead{USB Freq.} & \colhead{Array Config.} & \colhead{uv--range} & \colhead{Flux Calib.} & \colhead{Passband Calib.} & \colhead{Gain Calib.} & \colhead{$\tau_{\rm 225 GHz}$}\tablenotemark{a} \\
\colhead{} & \colhead{(YYYY-MM-DD)} & \colhead{(GHz)} & \colhead{(GHz)} & \colhead{} & \colhead{(k$\lambda$)} & \colhead{} & \colhead{} & \colhead{} & \colhead{} 
}
\startdata 
\multicolumn{9} {c} {Project:2024A-A004, PI: Chia-Ying Chung} \\
230 GHz–1 & 2024-10-19 & 193--205 & 213--225 & COM & 19--62 & Uranus & 3C84 & 0418+380, 0510+180 & $\sim$0.1 \\
          &            & 225--234 & 242--254 &     &      &        &      & \\
230 GHz–2 & 2024-11-03 & 193--205 & 213--225 & COM & 9--62 & Uranus & 3C84 & 0418+380, 0510+180 & $\sim$0.25\\
          &            & 225--234 & 242--254 &     &      &        &      & \\
\hline\multicolumn{9} {c} {Project:2024B-A001, PI: Chia-Ying Chung} \\
345 GHz-1  & 2024-12-01 & 331--343 & 351--363 & COM & 13--105  & Uranus, Callisto & 3C84 & V892~Tau, IC~2087~IR & $\sim$0.04 \\
          &            & 391.5--403.5 & 411.5--423.5 &     &      &        &      & \\
345 GHz-2  & 2024-12-24 & 331--343 & 351--363 & COM & 15--106  & Callisto & 3C84 & 3C84 & $\sim$0.05\\
          &            & 391.5--403.5 & 411.5--423.5 &     &      &        &      & \\
\enddata
\tablenotetext{a}{The atmospheric optical depth measured at 225 GHz.}
\end{deluxetable*}

The high dust masses and $\lesssim$100 $\mu$m maximum grain sizes in the protoplanetary disks may be concordantly explained if the dust coagulation is limited by the bouncing, fragmentation, or inward migration barriers (\citealt{Birnstiel2024ARA&A..62..157B}).
Such limitations make it hard to consume the dust mass budget in the formation of planetesimals and planets.

Since the sample of \citet{Chung2024ApJS..273...29C} is relatively bright, and may be spatially extended according to the known correlation between millimeter-luminosity and disk radius (\citealt{Andrews2018ApJ...869L..41A,Tripathi2017ApJ...845...44T,Hendler2020ApJ...895..126H}). The $\alpha_{\rm 200-400}$ may be relatively high if these extended objects tend to be associated with optically thin and spatially extended {\it disk halos}.
Whether or not $\alpha_{\rm 200-400}$ is in general $\lesssim$2.0 in the inner optically thick disk cores, and whether or not $a_{\rm max}$ is in general $\lesssim$100 $\mu$m, remain uncertain.
It is also unclear whether the fainter Class II objects represent those that have lost a significant fraction of the initial dust mass budget and thus may become optically thinner. 
To address these questions, it is necessary to extend the survey to Class II objects that are fainter at (sub)millimeter bands.

In this project, we focus on the Taurus-Auriga region so that the results can be compared with the survey of \citet{Chung2024ApJS..273...29C} without worrying about any potential systematic difference between the PPDs in different star-forming regions.
From \citet{Andrews2005ApJ...631.1134A} and \citet{Akeson_2019}, we selected 9 sources (Table \ref{tab:sources}) of which the 225 GHz flux densities were expected to be $\sim$10 mJy ($\sim$18 and 32 mJy at 300 and 400 GHz, respectively). 

Using the SMA, \citet{Andrews2013ApJ...771..129A} has detected 7 of these sources at $\sim$231 and $\sim$337 GHz.
However, in \citet{Andrews2013ApJ...771..129A}, the signal-to-noise ratios (SNR) of some of the objects were poor at at 337 GHz (e.g., CIDA~1, FZ~Tau, KPNO~10).
In addition, the 337 GHz flux density of FX~Tau published in \citet{Andrews2013ApJ...771..129A} was significantly offset from the independent measurement published in \citet{Akeson2014ApJ...784...62A}.
These led to uncertainties in the spectral indices ($\alpha$). 

To improve the constraints on $\alpha$, this project took advantage of the upgraded capability of the SMA to perform deeper observations, with better sampling in the spectral domain. 
Our observations are described in Section \ref{sec:observation}.
Details of how we calibrated the data and measured flux densities are introduced in Section \ref{sec:reduction}.
The results are presented in Section \ref{sec:results}.
Section \ref{sec:discussion} discusses our explanation for the present observational results and those of the observations of the brighter Taurus-Auriga disks presented in \citet{Chung2024ApJS..273...29C}, and the scientific implications.  
Section \ref{sec:conclusion} is a brief summary of our findings.
Our new SMA observations additionally included two bright and spatially compact Class II objects, IC~2087~IR and V892~Tau, that were observed by \citet{Chung2024ApJS..273...29C}, which allowed calibrating the systematic difference between our absolute flux scales and those of \citet{Chung2024ApJS..273...29C}.
This is introduced in Appendix \ref{appdx:flux}.
Appendix \ref{appdx:posterior} provides two examples of corner plots produced in the fittings of spectral indices.
Appendix \ref{appdx:radius} describes how we estimated the disk radii, which will be used in our statistical analyses.

\section{Observations}\label{sec:observation}
\subsection{Target sources}\label{sub:source}

To extend the SMA survey of \citet{Chung2024ApJS..273...29C} to objects that are fainter at (sub)millimeter bands, from  \citet{Andrews2005ApJ...631.1134A} and \citet{Akeson_2019}, we selected the following 10 objects in the Taurus-Auriga region,  CIDA-1, CX~Tau, FX~Tau, FZ~Tau, IT~Tau A/B, KPNO~10, V410~X-ray~2, V807~Tau, and 04301+2608 (Table \ref{tab:sources}), of which the 225 GHz flux densities were expected to be $\sim$10 mJy or lower.
The binary sources IT~Tau~A and IT~Tau~B can be covered within one field-of-view of the SMA observations.
Depending on the coordinates, the selected sources can be separated into two groups (Group-1: V410~X-ray~2, CIDA-1, CX~Tau, KPNO~10; Group-2: FZ~Tau, V807~Tau, 04301+2608, FX~Tau, IT~Tau), which requires different complex gain calibrators (see Section \ref{sub:obs}).

\subsection{SMA observations}\label{sub:obs}

We have performed the SMA observations in late 2024 (Project codes: 2024A-A004, 2024B-A001; PI: Chia-Ying Chung), which are summarized in Table \ref{tab:obs_summary}.
We used the dual-receiver mode. 
In this mode, each of the two receivers observed a single polarization.
The single polarization signal taken from each receiver was mixed with a local oscillator (LO) that can be independently tuned to produce signals in the upper and lower sidebands (USB and LSB). 
They were sampled by the SMA Wideband Astronomical ROACH2 Machine (SWARM; \citealt{Primiani2016JAI.....541006P}) correlator, which simultaneously covers the $\pm$4--16 GHz intermediate frequency (IF) ranges in the USB and LSB\footnote{The sensitivity provided by these SMA capabilities can be estimated using the online estimator \url{http://sma1.sma.hawaii.edu/beamcalc.html}.}. 

In each track (i.e., night) of our observations, the two independent LOs were tuned to different frequencies to obtain extended frequency coverages (Table \ref{tab:obs_summary}). 
There were four tracks of observations in total: the observations on 2024 October 19 and November 03 (hereafter tracks 230~GHz-1 and 230~GHz-2) used the receiver tunings with the LO at 209 GHz and 238 GHz, respectively; the observations on 2024 December 01 and 24 (hereafter tracks 345~GHz-1 and 345~GHz-2) used the receiver tunings with the LO at 347 GHz and 407.5 GHz, respectively.

There were 5 available antennae during the observations of track 230~GHz-1, and 6 available antennae during the observations of the remaining three tracks.
Data taken from the 238 GHz LO of track 230~GHz-1 and from the 407.5 GHz LO of track 345~GHz-1 were flagged during data reduction stages (see below) due to poor receiver responses. 

Our observations tracked the bright quasar, 3C84, for $\sim$30 minutes at the beginning of each track, and then observed it for 1.5 minutes every 30$\sim$60 minutes, to allow solving elevation-dependent passband amplitude and phase calibration solutions.
We observed the Solar system objects (Uranus, Callisto) at the beginning of each track of observations for absolute flux calibrations.

The target source loops are organized in the following way.
In each cycle, we either observed the target sources in Group-1 or Group-2.
For sources in Group-1 (Table \ref{tab:sources}), we sandwiched the scans on target sources with the observations on the distant quasars 0418$+$380 and the millimeter-bright Class~II disk V892\,Tau (\citealt{Chung2024ApJS..273...29C}); for sources in Group-2,  we sandwiched the scans on target sources with the observations on the distant quasars 0510$+$180 and the millimeter-bright Class~II disk IC~2087~IR (\citealt{Chung2024ApJS..273...29C}).
The target source loops in our observations have a $\sim$15 minutes cycle time (including the integration time on target sources, the integration time in the two scans on the complex gain calibrator and the two scans on the millimeter-bright Class~II disk that sandwiched the scans on target sources, and slewing time).
The reason for inserting the scans on V892~Tau and IC~2087~IR was due to the large angular separations between 0418$+$380, 0510$+180$, and the target sources (see Figure 24 in \citealt{Chung2024ApJS..273...29C}), and because 0418$+$380 and 0510$+180$ might only be detected at low signal-to-noise ratios at $\sim$400 GHz frequencies. 
Both can lead to noticeable residual gain phase errors at $>$300 GHz frequencies, which can be removed by applying the gain-phase self-calibration solutions derived from V892\,Tau and IC\,2087\,IR.
In addition, we can estimate the absolute flux errors by cross referencing the previous flux density measurements reported in \citet{Chung2024ApJS..273...29C}.

\begin{deluxetable}{lllll}[h]
\tabletypesize{\footnotesize}
\tablecolumns{5}
\tablewidth{0pt}
\tablecaption{Achieved rms noise levels \label{tab:rms}}
\tablehead{ 
\colhead{Source} & \colhead{$\delta_{\rm 209 GHz}$} & \colhead{$\delta_{\rm 238 GHz}$} & \colhead{$\delta_{\rm 347 GHz}$} & \colhead{$\delta_{\rm 407.5 GHz}$} \\
 & (mJy) & (mJy) & (mJy) & (mJy)
}
\startdata 
CIDA-1 & 1.7 & 1.7 & 2.1 & 12.8 \\
FZ Tau & 1.7 & 2.4 & 3.0 & 9.0 \\
V410 X-ray 2 & 1.9 & 1.9 & 2.8 & 10.9 \\
CX Tau & 1.1 & 1.3 & 1.8 & 7.9 \\
V807 Tau & 1.2 & 1.5 & 2.0 & 8.6 \\
FX Tau & 1.0 & 1.0 & 1.4 & 5.9 \\
IT Tau & 0.9 & 1.0 & 1.3 & 4.4 \\
KPNO 10 & 0.8 & 1.0 & 1.4 & 6.0 \\
04301+2608 & 0.8 & 1.3 & 1.3 & 5.4 \\
\enddata
\tablecomments{Root-mean-square noises ($\delta$) measured from the residual images produced by jointly imaging the upper and lower sidebands. From the second to the fifth columns, the subscripts in the column heads are the local oscillator frequencies (Section \ref{sub:obs}).
}
\end{deluxetable}

\section{Data Reduction}\label{sec:reduction}

\subsection{Routine calibration}\label{sub:calibration}

\begin{figure*}
\hspace{-1.6cm}
  \begin{tabular}{ cccc }
  \includegraphics[width=4.5cm]{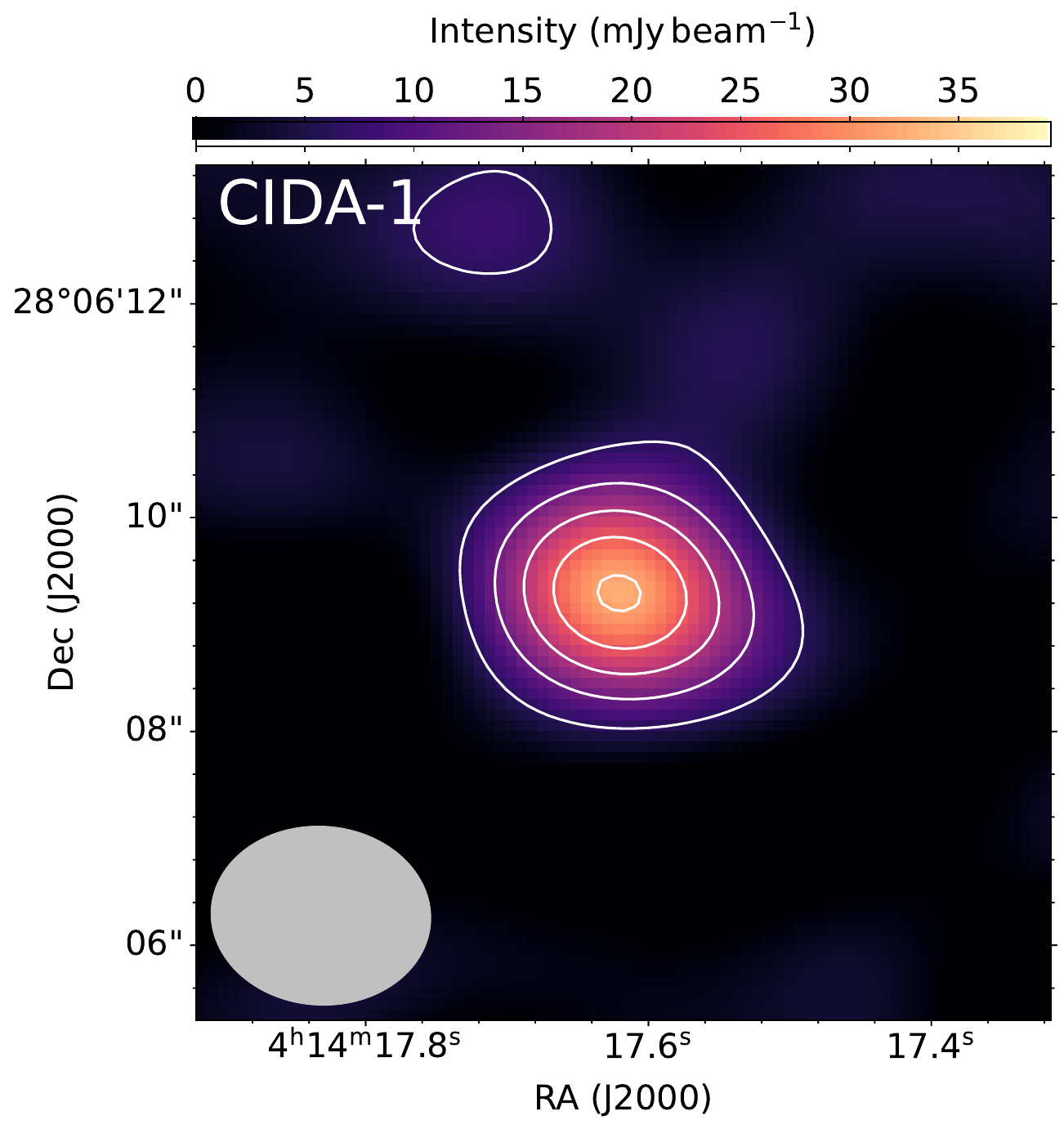} &
  \includegraphics[width=4.5cm]{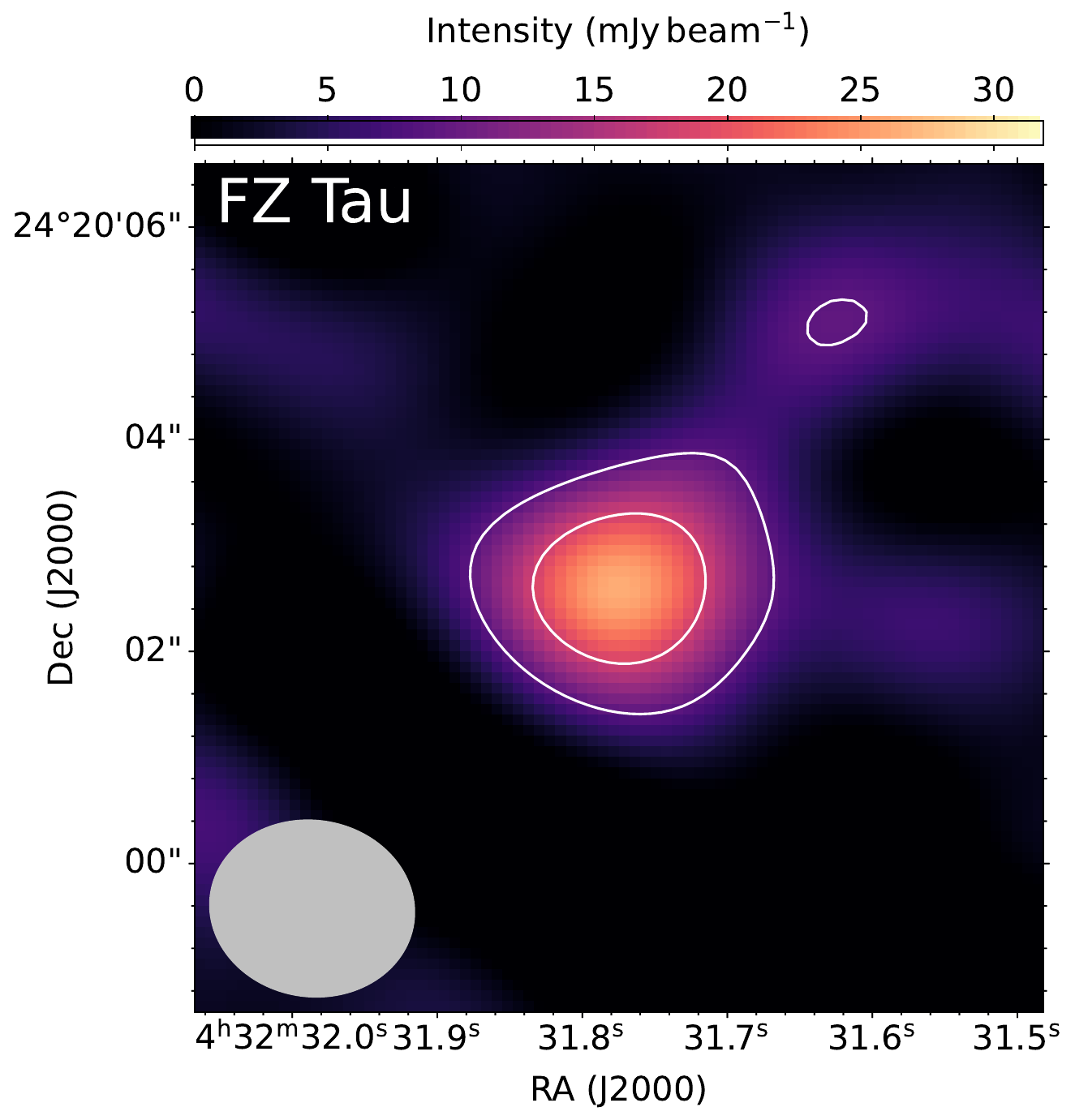} &
  \includegraphics[width=4.5cm]{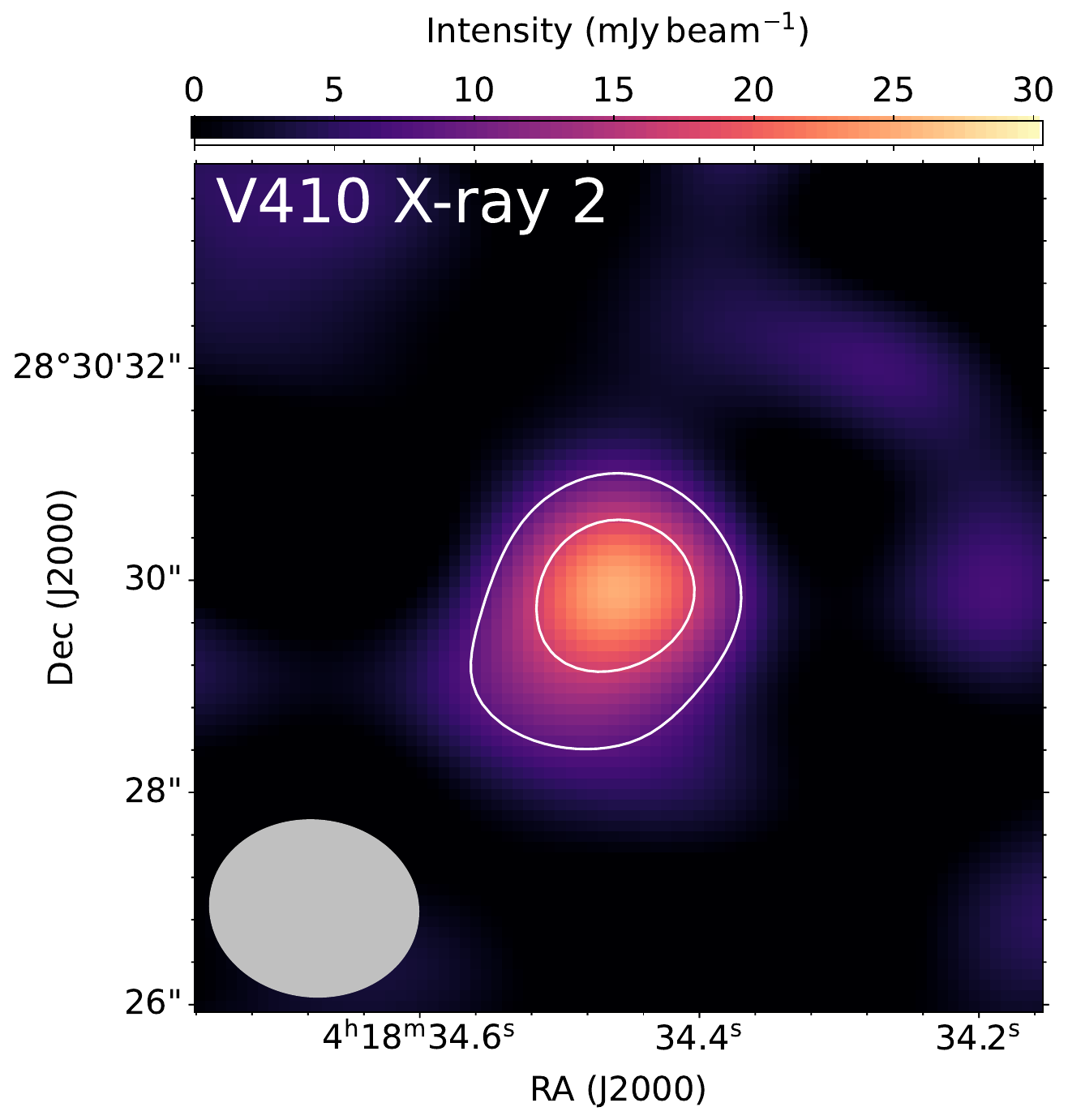} &
  \includegraphics[width=4.5cm]{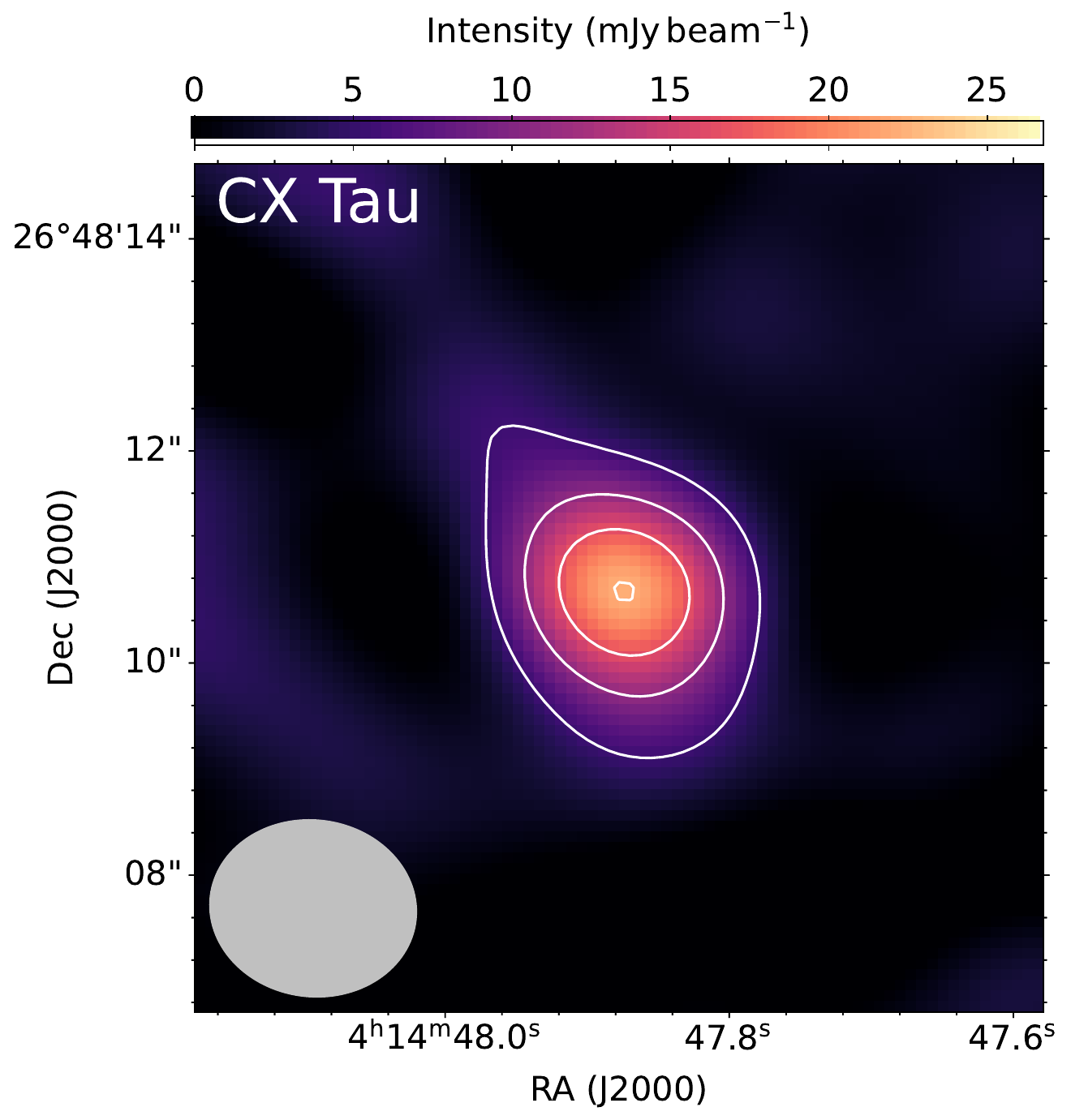}  \\
  \includegraphics[width=4.5cm]{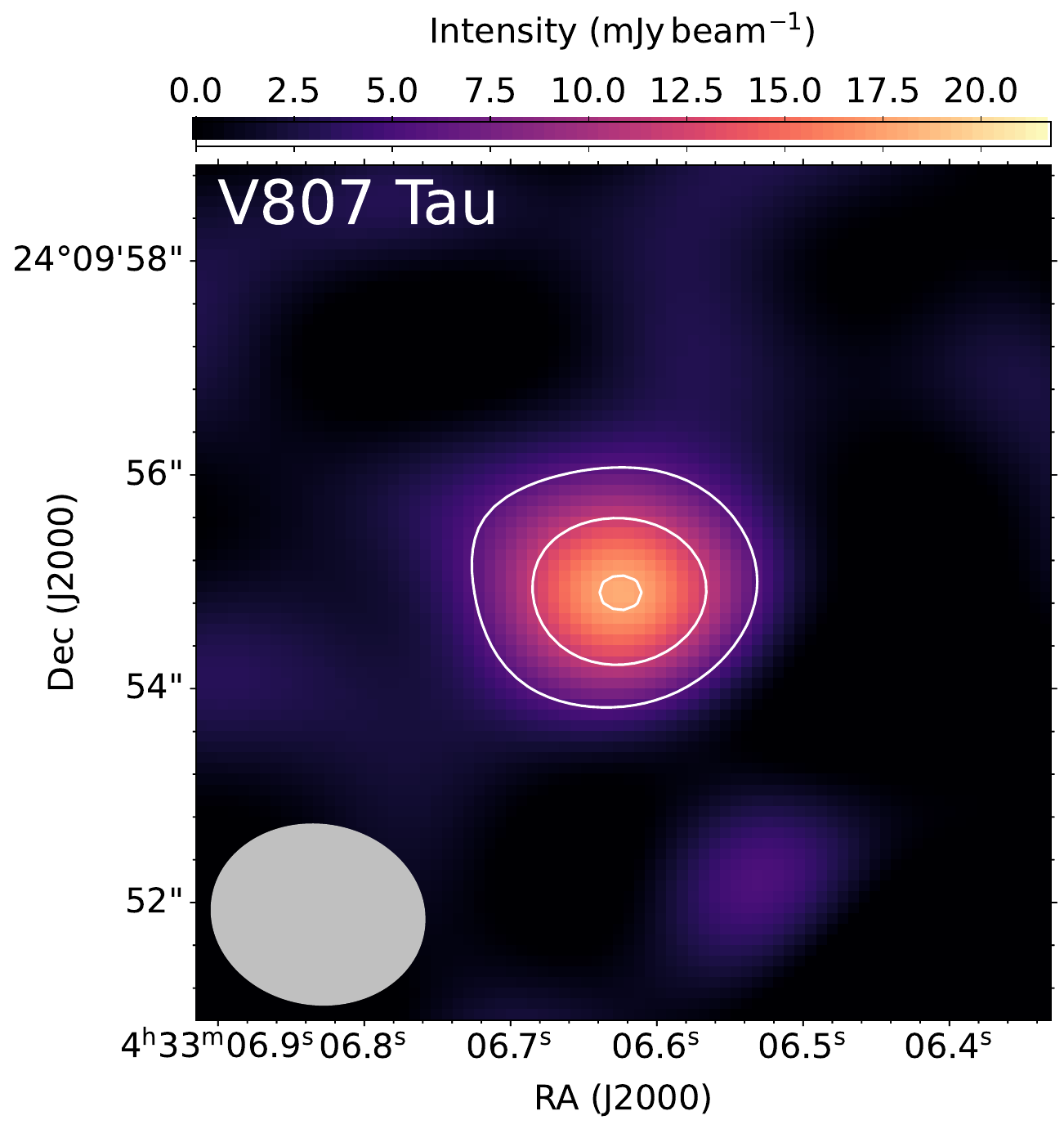} &
  \includegraphics[width=4.5cm]{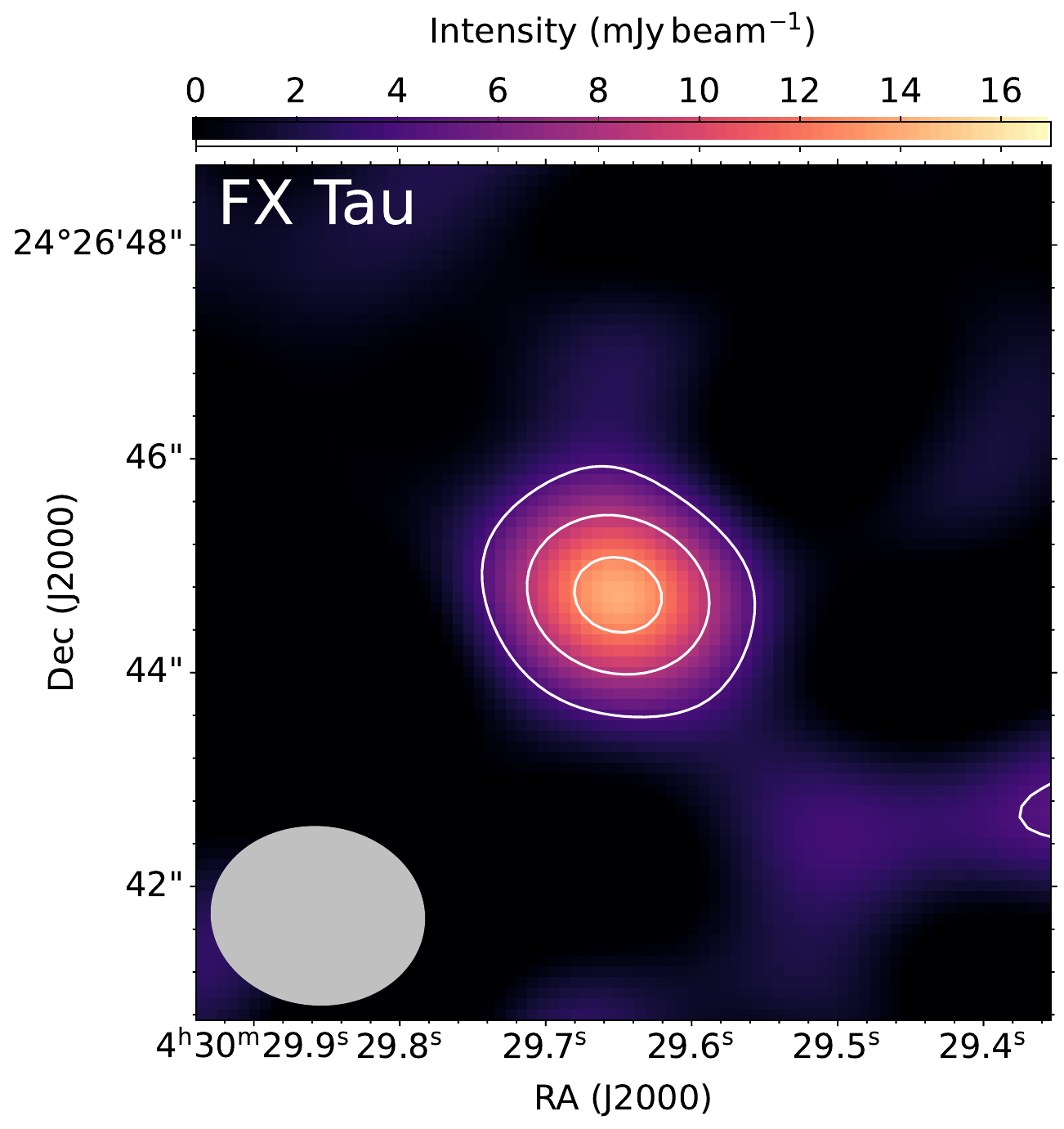} &
  \includegraphics[width=4.5cm]{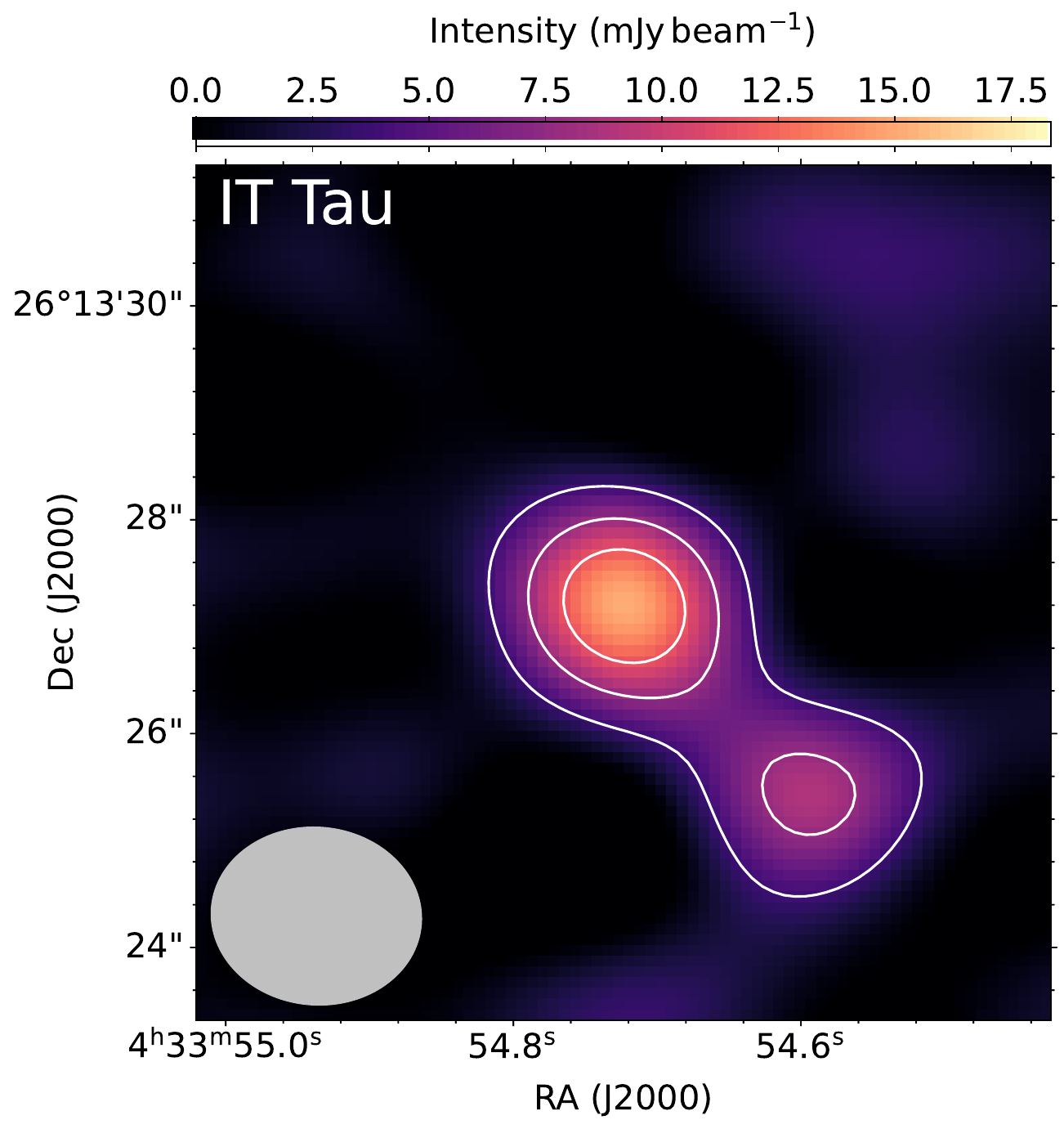} &
  \includegraphics[width=4.5cm]{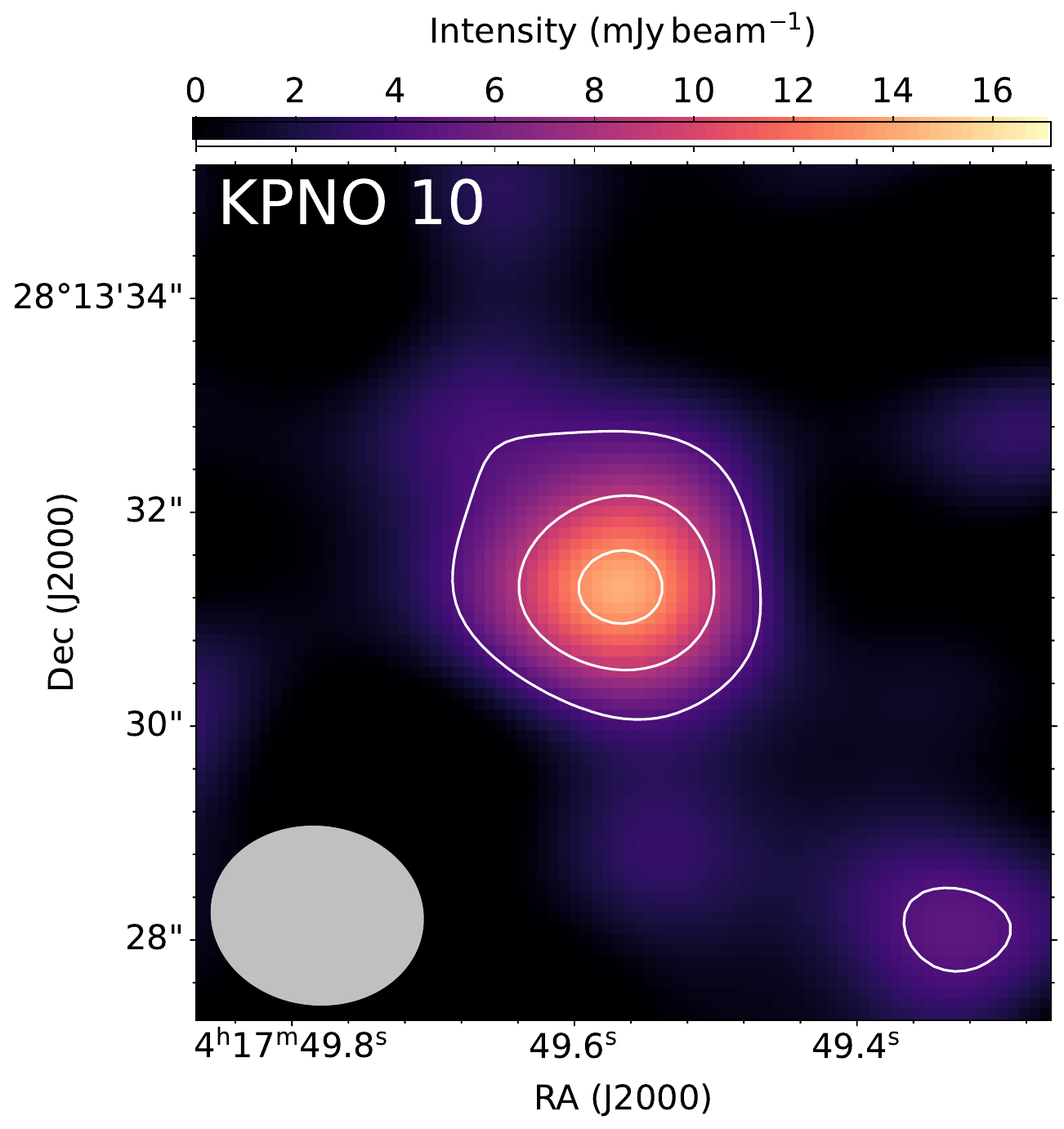} \\
  \includegraphics[width=4.5cm]{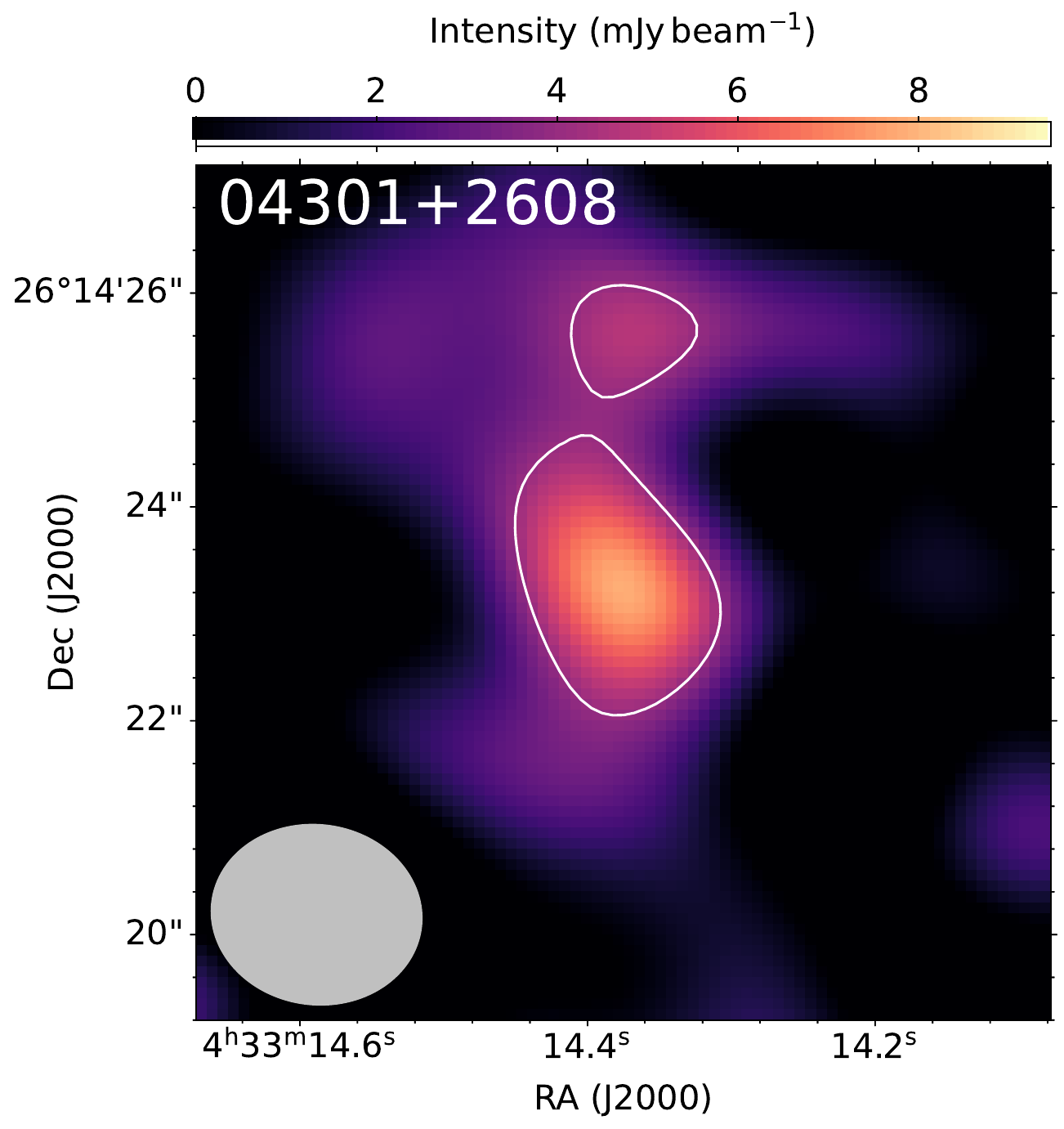} & & & \\
  \end{tabular}
  \caption{Images of the selected sources (Table \ref{tab:sources}) at 347 GHz. Each panel shows the 8$''$-wide square region centering at the peak of the target source. The IT~Tau panel covers the brighter source IT~Tau~A and the fainter companion IT~Tau~B southwest of it. Contours start at 3-$\sigma$, with 3-$\sigma$ intervals. For the sake of revealing the original observational results, when producing these images, we did not apply the additional multiplicative absolute flux calibration factors introduced in Section \ref{sub:flux density}.}\label{fig:rx400}
\end{figure*}

We performed basic data calibrations interactively using the MIR IDL (\citealt{Qi2003cdsf.conf..393Q}) software package. 
For each track of observations, we first visually inspected the raw complex visibilities to flag the data that are subject to issues like low response or phase jumps, etc.
In addition, we visually inspected the spectra of Uranus, Callisto, and 3C84, to ensure that there were no strong emission lines.
Afterwards, we applied the system temperature ($T_{\rm sys}$) measurements.

\begin{figure*}
\hspace{-2.5cm}
  \begin{tabular}{ cccc }
  \includegraphics[width=4.9cm]{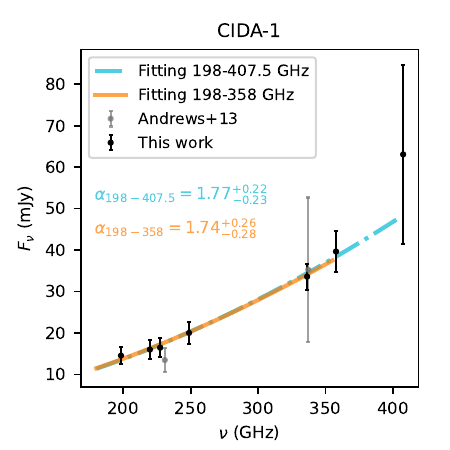} &
  \includegraphics[width=4.9cm]{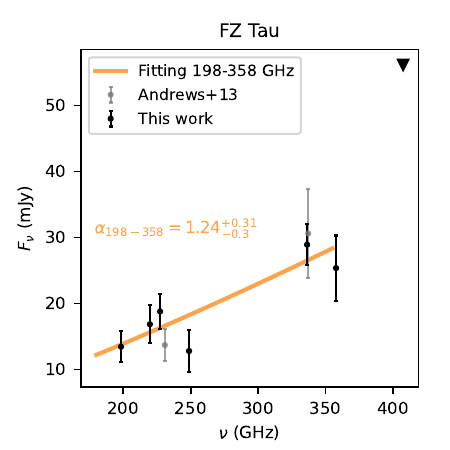} &
  \includegraphics[width=4.9cm]{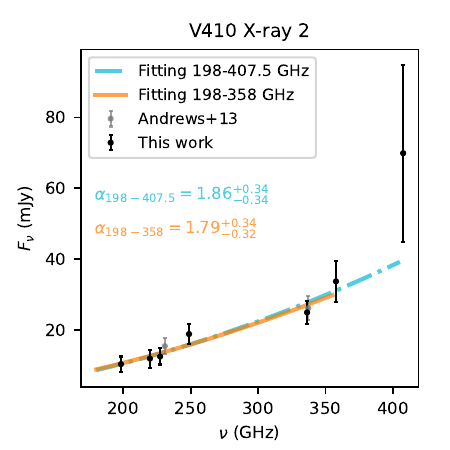} &
  \includegraphics[width=4.9cm]{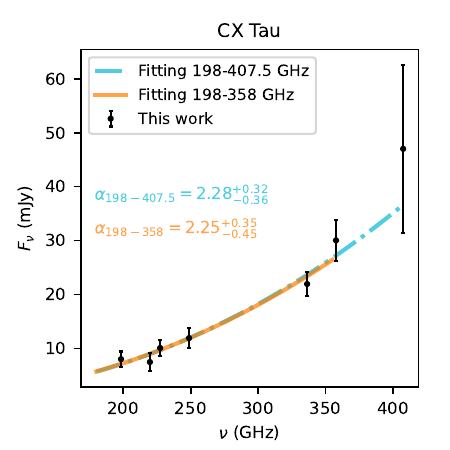}  \\
  \includegraphics[width=4.9cm]{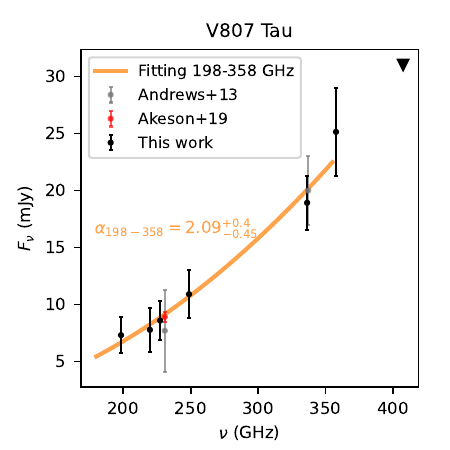} &
  \includegraphics[width=4.9cm]{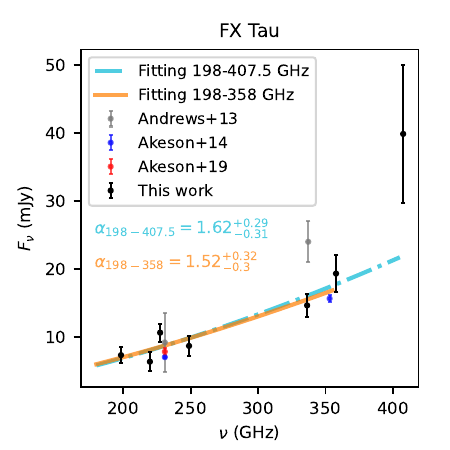} &
  \includegraphics[width=4.9cm]{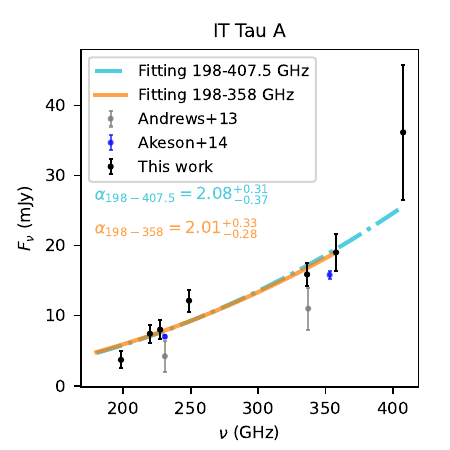} &
  \includegraphics[width=4.9cm]{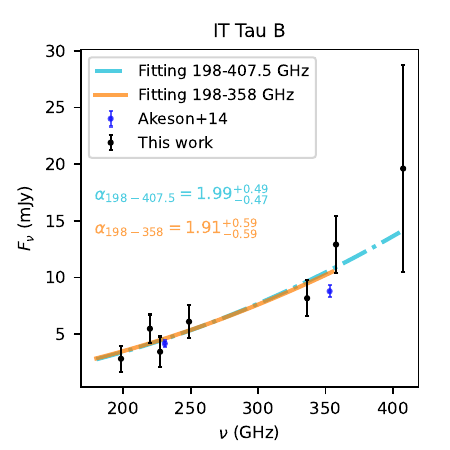} \\
  \includegraphics[width=4.9cm]{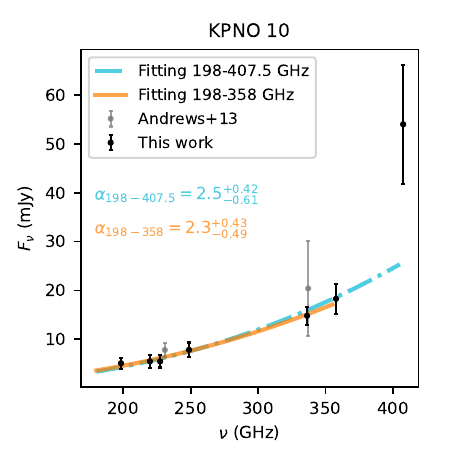} &
  \includegraphics[width=4.9cm]{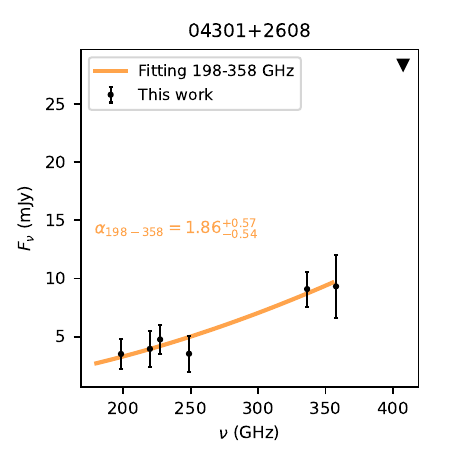} & &  \\
  \end{tabular}
  \caption{Flux densities of the selected Class II disks in the Taurus-Auriga region. The vertical error bars are $\pm$1$\sigma$ uncertainties. Upside-down triangles show 3-$\sigma$ upper limits.}
  \label{fig:Data}
\end{figure*}

For the observations of tracks 230~GHz-1 and 230~GHz-2, we derived the passband solutions based on averaging the spectra of 3C84 from all scans on it.
To mitigate the systematics caused by using spectrally unresolved $T_{\rm sys}$, which is not negligible at $\sim$300--400 GHz due to significant frequency variation of atmospheric transmission (\citealt{Chung2024ApJS..273...29C}), for the observations of tracks 345~GHz-1 and 345~GHz-2, we derived elevation ({\it El}) dependent passband solutions for the three elevation ranges, {\it El}$<$30$^{\circ}$, 30$^{\circ}<${\it El}$<$50$^{\circ}$, 50$^{\circ}<${\it El}. 
In a night of observations in December, 3C84 stayed in 30$^{\circ}$--50$^{\circ}$ elevation for $\sim$4 hours, and stayed in $>$50$^{\circ}$ elevation for $\sim$5 hours.
Since we observed 3C84 for 1.5 minutes every 30--60 minutes, our strategy of deriving {\it El}-dependent passband solution used approximately the same number of scans on 3C84 for each of the the 30$^{\circ}<${\it El}$<$50$^{\circ}$ and 50$^{\circ}<${\it El} elevation ranges. 
In each track, for each of these two elevation ranges, the time on 3C84 was approximately 10 minutes. 
Assuming the precipitable water vapor PWV$\sim$1 mm, which was the weather condition during our observations in 2024 December, a $\sim$10 minutes on-source time, at the $\sim$345 GHz observing frequency, for each pair of antennae (i.e., for each baseline), the expected root-mean-square (rms) noise in the velocity width of $\sim$100 km\,s$^{-1}$ is $\sim$300 mJy, which can detect 3C84 ($\sim$10 Jy) at 30-$\sigma$.
This is the SNR of the baseline-based passband solutions, which is lower than that of the antenna-based solutions we adopted.
In our experience, the passband calibration solutions with such an SNR are optimal for the scientific purpose of continuum observations.
Passband solutions derived in smaller elevation bins may be subject to poor SNR.
We remark that the estimate of rms noise is only inversely proportional to the square root of the exact on-source time. 
Finally, since we tracked 3C84 for 30 minutes at the beginning of each track of observations, when 3C84 was at elevation $<30^{\circ}$, the passband solutions for such low elevations have adequate SNR.

We followed the procedures described in the following Sections \ref{subsub:230cal} and \ref{subsub:345cal} to perform the first iteration of absolute flux and complex gain calibrations.
After applying all calibration solutions, we exported the calibrated data to the MIRIAD (\citealt{Sault1995ASPC...77..433S}) format for further processing.

\subsubsection{Absolute flux and gain calibration for tracks 230~GHz-1 and 230~GHz-2}\label{subsub:230cal}

We derived the absolute flux calibration solutions for each sideband based on the scans on Uranus.
In the track 230\,GHz-1 in all baselines, we found that the visibility amplitudes of the $T_{\rm sys}$-applied and passband-calibrated 3C84 data present time variations of up to $\sim$30\%.
We attribute this to the hour-angle and elevation dependent pointing errors. 
Similar effects were seen in the observations of track, 230\,GHz-2, in the baselines associated with antennae 2 and 8.
In the track, 230\,GHz-1, Uranus was located very close to 3C84 and our target sources.
We could accurately derive the absolute flux of 3C84, by referencing the observations on 3C84 and Uranus that were taken at proximate time (integration number 100--250).
We could then use 3C84 as the secondary absolute flux reference, which factored out the effects of hour-angle and elevation dependent pointing errors in the absolute flux calibration.
For the track, 230\,GHz-2, we excluded antennae 2 and 8 when performing absolute flux calibration to ensure that the absolute flux scales are not biased by any issues mentioned above or by technical issues that we have not understood.

We adopted the complex gain calibration solutions derived based on the observations on 0418+380 and 0510+180.
We also derived the complex gain calibration solutions based on the observations on IC~2087~IR and V892~Tau, which were sources of $\sim$200 mJy and $\sim$300 mJy in the frequency ranges covered by these two tracks of observations (Appendix \ref{appdx:flux}).
The solutions presented large scatter due to the limited SNR.
Therefore, we did not adopt them.

\subsubsection{Absolute flux and gain calibration for tracks 345~GHz-1 and 345~GHz-2}\label{subsub:345cal}

We derived the absolute flux calibration solutions for each sideband based on the scans on Callisto.
We derived the complex gain calibration solutions for the observations at the 347 GHz LO frequency based on the observations on IC~2087~IR and V892~Tau, which were $\sim$700 mJy and $\sim$900 mJy sources in the frequency ranges covered by these two tracks of observations (Appendix \ref{appdx:flux}).
The flux densities of these two unresolved Class II disks are comparable with those of 0418$+$380 and 0510$+$180.
In terms of SNR, the quality of the calibration solutions derived from these two unresolved Class II disks is comparable to the calibration solutions derived from 0418$+$380 and 0510$+$180.
Because IC~2087~IR and V892~Tau are located much closer to the selected target sources, adopting the calibration solutions derived from these two unresolved disks mitigates the residual phase errors caused by the direction dependence on atmospheric phase fluctuations.

Absolute fluxes of IC~2087~IR and V892~Tau at the upper and lower sidebands were obtained by comparing the visibility amplitudes with those of Callisto.
We derived the complex gain calibration solutions for the observations at the 407.5 GHz LO based on the observations on 3C84, since other calibration sources were not detected with high enough SNR.
For example, in the 345~GHz-1 track (Table \ref{tab:obs_summary}), over the frequency range of 391.5--403.5 GHz, at elevations of 55$^{\circ}$--60$^{\circ}$, in a scan of $\sim$1.5 minutes, the SNRs of the antenna-based gain calibration solutions derived from 3C84, 0418+380, V892~Tau, 0510+180, and IC~2087~IR are 18, 6, 2, 3, and 4, respectively (the SNRs are lower in the observations at lower elevations due to the higher airmass in the line of sight).





\subsection{Imaging}\label{sub:imaging}
We produced naturally weighted (\citealt{Briggs1999ASPC..180..127B}) images using the MIRIAD software package (\citealt{Sault1995ASPC...77..433S}).
For the observations at each of the four LOs (Section \ref{sub:obs}), we jointly imaged the data taken from the upper and lower sidebands.
We first created dirty images using the \texttt{invert} task, from which we identified the peaks of the emission.
We then deconvolved the dirty images using the \texttt{clean} task for 100 iterations, restricted in the 5$''$-wide box regions around the peaks and without setting cutoff levels.
This yielded the preliminary residual images.
We measured the rms noise levels from these preliminary residual images.
To produce the final images, we re-deconvolved the dirty images by setting the cutoff levels to 1.5 times the rms noises. 

For individual sources,  the rms noises in the final images are listed in Table \ref{tab:rms}.
The typically achieved synthesized beams at the 209, 238, 347, and 407.5 GHz LOs are (3\farcs5$\times$2\farcs5; P.A.$=-$90$^{\circ}$), (2\farcs5$\times$2\farcs5; P.A.$=$80$^{\circ}$), (2\farcs0$\times$1\farcs5; P.A.$=$90$^{\circ}$), and (2\farcs5$\times$1\farcs0; P.A.$=-$80$^{\circ}$)\footnote{The angular scales are the FWHM along the major and minor axes of the synthesized beam.}, respectively.

\subsection{Measuring flux densities}\label{sub:flux density}
The angular diameters of the selected target sources are in the range of 170--430 mas (see Appendix \ref{appdx:radius}), which are too small to be resolved by our new SMA observations (Section \ref{sub:obs}).
In addition, except for the 2\farcs4 separated binaries IT~Tau~A and IT~Tau~B, each of the selected objects is the only object detected in the field of view of the SMA observations for it. 

We measured the flux densities from fits to the visibility data, and fits to the images.
For the 209 GHz, 238 GHz, and 347 GHz LOs (Section \ref{sub:obs}), for each object, we derived the flux densities in each sideband based on fitting the complex visibilities as a two-dimensional (2D) Gaussian source (the resolved binary components IT~Tau~A and IT~Tau~B were treated as independent Gaussian sources).
For the 407.5 GHz LO, we fitted the visibility data taken from both sidebands jointly for a better SNR.
This yielded a total of 7 independent samples in the 200--420 GHz frequency range.
The free parameters are flux density (flux), x-offset (dx), y-offset (dy), and Gaussian full width at half maximum (FWHM).
The detailed procedures (including prior functions) is identical to those outlined in Section 3.4 of \citet{Chung2024ApJS..273...29C}.

When the observations are subject to phase noise, the data taken from relatively long baselines may be impacted more. 
The degraded flux densities at long baselines mimic the visibilities of a spatially resolved source. 
In such cases, fitting point sources may lead to underestimates of flux densities; using a Gaussian model alleviates this problem.
We also fitted the visibility data as point sources. 
The results are not distinguishable from the 2D Gaussian fits.

We first measured the flux densities of IC~2087~IR and V892~Tau, independently from all four tracks of observations (Table \ref{tab:obs_summary}).
We then compared these flux density measurements with those reported in \citet{Chung2024ApJS..273...29C} to derive the flux rescaling factors (see Appendix \ref{appdx:flux}).
After applying the flux rescaling factors to the visibilities of all objects, we measured the flux densities by jointly fitting the visibilities taken from tracks 230 GHz-1 and 230 GHz-2, and jointly fitting those taken from tracks 345 GHz-1 and 345 GHz-2.
For tracks 230 GHz-1 and 230 GHz-2, this strategy effectively removed the residual absolute flux errors caused by hour-angle and elevation dependent pointing errors (Section \ref{subsub:230cal}).

\subsection{SED fitting and spectral index}\label{sub:SED fitting}

To derive spectral indices, we employed the {\tt emcee} package (\citealt{Foreman-Mackey2013PASP}) to fit the observed SED of each target source by a single power law using the Markov Chain Monte Carlo (MCMC) method.
The free parameters are the flux density at 200 GHz wavelength ($F_{\rm 200 GHz}$) and the spectral index ($\alpha$).
We note that based on similar SMA observations of 47 Class II disks, \citet{Chung2024ApJS..273...29C} did not detect frequency variation of spectral indices over the 200--400 GHz frequency range. 
Therefore, we think that a single power-law is a good approximation of our presently observed spectra (Figure \ref{fig:Data}).

The prior functions of $F_{\rm 200 GHz}$ and $\alpha$ are uniform in the ranges of $-10^{-10}$--$10^{10}$ and 0--5, respectively. 
Using these flat prior functions, essentially, the MCMC walkers were sampling the likelihood functions inside the defined parameter ranges. 
Compared to the fitting methods that directly maximize the likelihood, the MCMC method is advantageous since it is not trapped by the local maxima of the likelihood. 

We adopted 60 walkers and ran 1500 steps, discarding the initial 500 steps as burnt-in steps. 
For each walker, the initial $F_{\rm 200 GHz}$ and $\alpha$ were Gaussian random samples around 20 mJy and 2.0, with standard deviations of 3.0 mJy and 1.0, respectively. 
The best-fit values of $F_{\rm 200 GHz}$ and $\alpha$ are the coordinates at the peak of the posterior probability. 
Fitting errors are defined as the differences between the best fit values and the 16th and 84th percentiles of the MCMC samples (\citealt{Foreman-Mackey2013PASP}).
Examples of corner plots are provided in Appendix \ref{appdx:posterior}.

The following discussion is mainly based on the spectral indices derived in the frequency range of 198--358 GHz (hereafter $\alpha_{\rm 198-358}$), since our flux density measurements in this frequency range achieved several times better SNR than the measurements at 407.5 GHz.
We included the spectral indices derived in the wider frequency range of 198--407.5 GHz (hereafter $\alpha_{\rm 198-407.5}$) for references.

This study aims at comparing the spectral indices derived for our selected objects with the spectral indices derived for the 47 brighter Class II objects in the Taurus-Auriga region \citep{Chung2024ApJS..273...29C}.
In case of systematic biases, we used our fitting procedure described above to re-fit these data, and confirmed that the derived $\alpha_{\rm 200-400 GHz}$ are consistent with those published in \citet{Chung2024ApJS..273...29C}.

\section{Results} \label{sec:results}

Except for the 407.5 GHz LO observations on FZ~Tau, V807~Tau, and 04301+2608, the observations detected the selected objects (Figure \ref{fig:rx400}, \ref{fig:Data}).
With the angular resolutions we achieved (Section \ref{sub:imaging}), the binary components IT~Tau~A and IT~Tau~B (Table \ref{tab:sources}) were spatially resolved (Figures \ref{fig:rx400}, \ref{fig:Data}).
In the following discussion, they are treated as two independent objects. 
We treated FX~Tau as one object, since we could not resolve its binary components. 
Similarly, we treated the triple system, V807~Tau (\citealt{Schaefer2012ApJ...756..120S}), as one object since we could not spatially resolve them. 
We did not detect the $\sim$16$''$ separated companion of FZ~Tau, FY~Tau.
Other sources appear as isolated sources in our SMA images (Figure \ref{fig:rx400}).
In summary, our observations on the 9 target sources detected the (sub)millimeter emission from 10 independent Class II objects. 

Figure \ref{fig:Data} shows the observed flux densities (Section \ref{sub:flux density}) and the spectral indices $\alpha_{\rm 198-358}$ and $\alpha_{\rm 198-407.5}$ of the 10 objects, which were derived based on fitting power-laws in the frequency ranges of [198, 358] and [198, 407.5] GHz, respectively (Section \ref{sub:SED fitting}).
Flux densities are summarized in Table \ref{tab:flux}.

Due to the high uncertainties of the 407.5 GHz flux densities (Figure \ref{fig:Data}), the measurements at 407.5 GHz have only a small
effect on the spectral indices fits. 
As a result, the derived values of $\alpha_{\rm 198-358}$ and $\alpha_{\rm 198-407.5}$ are very similar.
A few objects (V410~X-ray~2, FX~Tau, IT~Tau~A, KPNO~10) show an excess of flux densities at 407.5 GHz with respect to the derived power-laws.
In \citet{Chung2024ApJS..273...29C}, such excesses at high-frequency were interpreted as dust thermal emission from spatially relatively extended, optically thin {\it disk halos}, which have relatively high spectral indices (e.g., 3--3.8).
In our cases, it can also {\bf be} because we over-corrected the 407.5 GHz flux densities in the post-processing (Appendix \ref{appdx:flux}).
These two possibilities are not mutually exclusive. 
We note that the possibility that we over-corrected the 407.5 GHz flux densities may imply that \citet{Chung2024ApJS..273...29C} systematically overestimated the flux densities at $>$400 GHz, which in turn led to over-estimates of $\alpha_{\rm 200-400}$ in that study.
If this is indeed the case, then the actual mean value of $\alpha_{\rm 200-400}$ will be lower than the 2.0$\pm$0.2 \citet{Chung2024ApJS..273...29C} reported. 
This needs to be checked by re-observing the sample of \citet{Chung2024ApJS..273...29C} at $>$400 GHz frequencies. 
The following discussion focuses on $\alpha_{\rm 198-358}$, which is more robust. 

\begin{figure}
    \hspace{-0.3cm}
    \includegraphics[width=9cm]{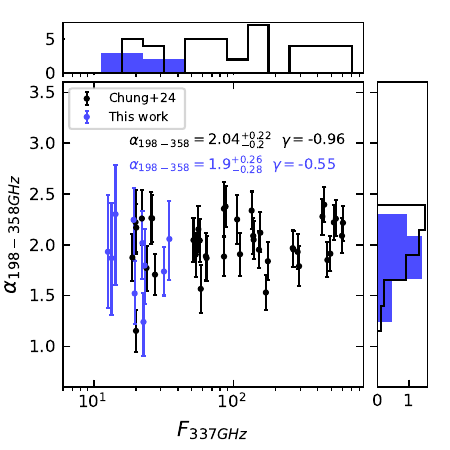}
    \caption{The 198--358 GHz spectral index ($\alpha_{\rm 198-358}$) versus 337 GHz flux density ($F_{\rm 337 GHz}$). The values of $F_{\rm 337 GHz}$ have been rescaled to a 140 pc distance, according to the distances of the target sources listed in Table \ref{tab:sources}. The histograms on the right have been normalized such that the area under each of them is 1.0.
    We excluded the extended disks, DL~Tau, CI~Tau, GM~Tau, AB~Aur, DM~Tau, AA~Tau, and GO~Tau from the sample of \citet{Chung2024ApJS..273...29C}.
    }
    \label{fig:Fvsalpha200400}
\end{figure}

Figure \ref{fig:Fvsalpha200400} displays the distribution of $\alpha_{\rm 198-358}$.
Among our 10 detected objects, the median and standard deviation of $\alpha_{\rm 198-358}$ are 1.9 and 0.3, respectively.
The distribution of $\alpha_{\rm 198-358}$ derived from these 10 objects is low-value-skewed, with a skewness\footnote{For a random variable $X$, the skewness is the third standard moment $\frac{E[(X-\mu)^3]}{\sigma^3}$, where $E[\cdot]$ denotes the expectation value, $\mu$ is the mean of the distribution, and $\sigma$ is the standard deviation.} of $-$0.55.
We compared the distribution of $\alpha_{\rm 198-358}$ from these 10 objects with the $\alpha_{\rm 198-358}$ derived from the sample of \citet{Chung2024ApJS..273...29C}, after excluding the aforementioned 7 extended objects which were spatially resolved and present systematically higher $\alpha_{\rm 200-400}$ values.
The two $\alpha_{\rm 198-358}$ distributions appear very similar, both with low-value-skewed distributions (Figure \ref{fig:Fvsalpha200400}).
We used the Kolmogorov–Smirnov (KS) test and the non-parametric  Mann–Whitney U test to verify the null hypothesis that these two $\alpha_{\rm 198-358}$ distributions are not distinct. 
We used the \texttt{ks\_2samp} and \texttt{mannwhitneyu} functions of the scipy package (\citealt{virtanen2020scipy}).
To understand the effect of the uncertainties of $\alpha_{\rm 198-358}$ in these tests, in each test, we evaluated the $p$ values of 5000 random realizations of the actual $\alpha_{\rm 198-358}$ values assuming that the errors of $\alpha_{\rm 198-358}$ followed Gaussian distributions.
In the KS test, the median value of $p$ is {\bf 0.25}; the fraction of $p<$0.05 is 0.17.
In the Mann–Whitney U test, the median value of $p$ is 0.29; the fraction of $p<$0.05 is 0.15.
The results of both tests did not reject the null hypothesis. 
The Pearson correlation coefficient of all the objects in Figure \ref{fig:Fvsalpha200400} is 0.24, with a p-value of 0.093.

\begin{figure}
    \hspace{-0.3cm}
    \includegraphics[width=9cm]{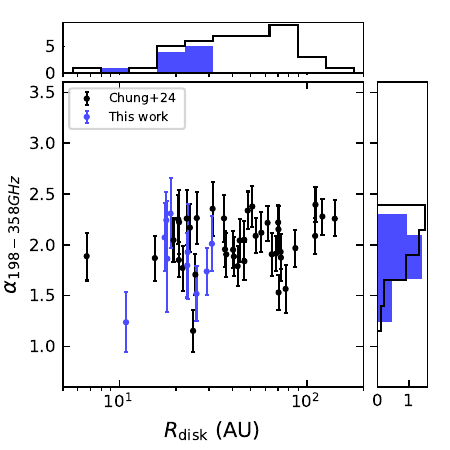}
    \caption{The 198--358 GHz spectral index ($\alpha_{\rm 198-358}$) versus disk radius ($R_{\rm 95\%}$). The histograms on the right have been normalized such that the area under each of them is 1.0.}
    \label{fig:Rvsalpha200400}
\end{figure}

Figure \ref{fig:Rvsalpha200400} shows the relations between the spectral indices and the disk radii ($R_{\rm 95\%}$), where $R_{\rm 95\%}$ denotes the radius of a circular region that encloses 95\% of the total flux density of a source (Appendix \ref{appdx:radius}).
From all the objects presented in this figure, we do not see a convincing correlation between $R_{\rm 95\%}$ and $\alpha_{\rm 198-358}$.
The Pearson correlation coefficient between these two quantities is 0.25, with a $p$ value of 0.09, consistent with our visual impression of no correlation.

\section{Discussion} \label{sec:discussion}
A simple explanation for our result of $\alpha_{\rm 198-358}=1.9\pm0.3$ is that the $\sim$200--400 GHz flux densities of these objects are dominated by optically thick dust thermal emission.
However, when there is some turbulence in the gaseous protoplanetary disks, the diffusion of dust particles can broaden the dusty structures (e.g., rings, see the discussion in the recent theoretical study of \citealt{Yang2025ApJ...989..176Y} and references therein) by several au, which may be comparable to the intrinsic widths of the pressure bumps that trap dust (e.g., \citealt{Dullemond2018ApJ...869L..46D}).
A dusty structure may not be optically thick everywhere in a frequency band. 
Therefore, when the observed spectral index of a bulk of a dusty structure is close to 2.0, it may be the result of mixing the emission from the optically thinner region that has $\alpha>$2.0, with some emission with $\alpha<$2.0.
Due to dust self-scattering (\citealt{Liu2019ApJ...877L..22L,Zhu2019ApJ...877L..18Z}), the $\alpha<$2.0 values occur in regions with high optical depths and when the dust albedo increases with frequency in the observed frequency range.
In our observations, this corresponds to a maximum dust grain size of $\sim$100--300 $\mu$m (\citealt{Liu2019ApJ...877L..22L,Zhu2019ApJ...877L..18Z}).

A more detailed analysis of the spectrum observed in a broader frequency range indicated that this is likely the case in one of our selected objects, CIDA~1, which did not show strong evidence of free-free emission in centimeter bands (\citealt{Hashimoto2023AJ....166..186H}).
Whether or not this is a general result remains uncertain. 

In some, but not necessarily all of the remaining selected objects, the low $\alpha_{\rm 198-358}$ values may also be attributed to the contamination of free-free emission. 
The previous JVLA observations resolved free-free emission in CX~Tau; the flux density and spectral index are $\sim$0.4 mJy and $\sim$0.7 at 33 GHz (\citealt{Curone2023A&A...677A.118C}).
The $\alpha_{\rm 198-358}=1.24\pm0.3$ value we resolved in FZ~Tau (Figure \ref{fig:Data}) might be explained by free-free emission, if it is not largely due to measurement errors.
\citet{Pietu2014A&A...564A..95P} reported that the  spectral index of FZ~Tau at the lower, $\sim$100--200 GHz frequency range is 1.6$\pm$0.5.
Due to the large uncertainties of these spectral indices, the nature of the (sub)millimeter emission in FZ~Tau remains uncertain. 
We note that although the $\sim$0\farcs3 synthesized beam achieved by \citet{Pietu2014A&A...564A..95P} is $\sim$10 times smaller ours (Section \ref{sub:imaging}), FZ~Tau remains spatially unresolved in their observations.
Therefore, missing short-spacing is not a concern in the comparison between their and our observations.

The free-free emission in protoplanetary disks may have $-$0.1--2.0 spectral indices (c.f. \citealt{Anglada2018}).
In some protoplanetary disks in the Taurus-Auriga region, free-free emission may dominate the flux density at low frequencies (\citealt{Chung2025ApJS..277...45C}), which makes the observed spectral indices lower at lower frequencies.
Among our selected target sources, the earlier JVLA survey (\citealt{Dzib2015ApJ...801...91D}) shows that the flux densities of V410~X-ray~2 at 4.5 GHz and 7.5 GHz are both $\sim$0.2 mJy, which are $\sim$50 times lower than the flux densities detected at $\sim$200 GHz (Table \ref{tab:flux}); the flux densities of V807~Tau at 4.5 GHz and 7.5 GHz are both $\sim$0.28 mJy, which are $\sim$30 times lower than the flux densities detected at $\sim$200 GHz (Table \ref{tab:flux}).
If V410~X-ray~2 and V807~Tau do not present large variabilities (c.f. \citealt{Dzib2015ApJ...801...91D,Liu2014ApJ...780..155L}) of free-free emission (i.e., if the free-free emission in these sources in 2024 is not considerably brighter than their free-free emission in the observing time epochs of \citealt{Dzib2015ApJ...801...91D}), then free-free emission may not explain the low $\alpha_{\rm 198-358}$ values in these two sources.
We do not find constraints on the flux densities of free-free emission in the other selected objects.

The observed spectral indices can be explained by high dust optical depths ($\tau\gtrsim$5), as suggested by \citet{Chung2024ApJS..273...29C}, although a limited range of dust temperatures across the sample could also contribute to the relatively uniform $\alpha$ values (see their Section 5.2.2). 
These effects are not mutually exclusive.

There is no clear statistical indication that the dust emission properties of our resolved 10 sources are different from the optically thick {\it disk cores} of the 47 (sub)millimeter-brighter Class II disks in the Taurus-Auriga region reported in \citet{Chung2024ApJS..273...29C}.
If the dust emission in the 10 selected Taurus-Auriga Class II objects and the 47 brighter Taurus-Auriga Class II disks reported in \citet{Chung2024ApJS..273...29C} are all optically thick at 200--400 GHz, then the (sub)millimeter observations only provide lower limits for their dust masses (\citealt{Hildebrand1983QJRAS..24..267H}). 
With the accretion rates of Class II objects (\citealt{Manara2023ASPC..534..539M}), Class II disks can typically be considered as passive disks, in which the dust temperature depends weakly on protostellar luminosity and other disk properties (\citealt{Chiang1997ApJ...490..368C}).
In this case, when a source is weak at (sub)millimeter wavelengths, it only indicates that the dusty disk of this source has a small projected area rather than having a small dust mass. 
The age-evolution of (sub)millimeter flux densities (e.g. \citealt{Ansdell2017AJ....153..240A,Williams2019ApJ...875L...9W}) may be explained either by dust mass dispersal or by the age-evolution of the radii of dusty disks (\citealt{Hendler2020ApJ...895..126H}).
Physically, the latter may be due to the inward drift of the grown dust or the radial trapping of the grown dust to geometrically narrow substructures  (\citealt{Testi2014prpl.conf..339T,Birnstiel2024ARA&A..62..157B}).
Finaly, it may be equally plausible that an initial distribution of disk sizes accounts for the distribution of (sub)millimeter flux densities, and there is no signature of evolution.

We cannot yet rule out the possibility that some of our 10 observed objects are optically thin at 200--400 GHz.
In such cases, the close to 2.0 values of $\alpha_{\rm 198-358}$ may be interpreted by dust grain growth that considerably lowered the dust (absorption) opacity spectral indices (e.g., $a_{\rm max}\gg$1 cm; c.f. \citealt{Testi2014prpl.conf..339T}).
This possibility can be tested by resolving the dust brightness temperature at 200--400 GHz with future high angular resolution observations, or by extending the flux density measurements to lower frequencies {\bf (\citealt{Sierra2021ApJS..257...14S})}. 

\section{Summary} \label{sec:conclusion}
Using the SMA, we measured the 198.0--407.5 GHz spectra for 10 Class II disks in the Taurua-Auriga region.
The 337 GHz flux densities of these objects are in the range of 8.2--34 mJy (Table \ref{tab:flux}).
We found that the median and standard deviation of the spectral index $\alpha_{\rm 198-358}$ of these 10 objects are 1.9 and 0.3, respectively.
The distribution of $\alpha_{\rm 198-358}$ derived from these 10 Class II disks is low-value-skewed.
To our measurement uncertainties, this $\alpha_{\rm 198-358}$ distribution cannot be distinguished from that derived from other 47 brighter Class II disks in the Taurua-Auriga region.
The tentative interpretation is that the 10 Class II disks selected for our present study and the compared 47 brighter Class II disks in the Taurua-Auriga region are optically thick in the (sub)millimeter bands, with optical depths $\tau\gtrsim$5. 
This can be confirmed by performing higher angular resolution observations or observations at lower frequency bands.

\begin{acknowledgments}
The Submillimeter Array is a joint project between the Smithsonian Astrophysical Observatory and the Academia Sinica Institute of Astronomy and Astrophysics and is funded by the Smithsonian Institution and the Academia Sinica (\citealt{Ho2004ApJ...616L...1H}).
We recognize that Maunakea is a culturally important site for the indigenous Hawaiian people; we are privileged to study the cosmos from its summit.
This paper makes use of the following ALMA data: ADS/JAO.ALMA\#2021.1.00854.S, \#2016.1.00715.S, \#2013.1.00105.S, \#2022.1.01302.S, and \#2016.1.01511.S. 
ALMA is a partnership of ESO (representing its member states), NSF (USA) and NINS (Japan), together with NRC (Canada), MOST and ASIAA (Taiwan), and KASI (Republic of Korea), in cooperation with the Republic of Chile. The Joint ALMA Observatory is operated by ESO, AUI/NRAO and NAOJ. 
C.Y.C., and H.B.L. are supported by the National Science and Technology Council (NSTC) of Taiwan (Grant Nos. 113-2112-M-110-022-MY3).
\end{acknowledgments}

\begin{contribution}

C.Y.C. prepared the proposal and the observing scripts, and performed visibility fittings.
H.B.L. performed data calibration and spectral index fittings.
S.M.A assisted visibiilty fittings.
M.A.G. assisted the planning of observations and absolute flux calibrations. 
All authors participated in the scientific discussion. 


\end{contribution}

%
\facilities{SMA, ALMA}

\software{
          {\tt astropy} \citep{astropy2022ApJ...935..167A},  
          {\tt Numpy} \citep{VanDerWalt2011}, 
          {\tt scipy} \citep{virtanen2020scipy},
          {\tt MIR IDL} \citep{Qi2003cdsf.conf..393Q},
          {\tt Miriad} \citep{Sault1995ASPC...77..433S},
          {\tt emcee} \citep{Foreman-Mackey2013PASP},
          }


\begin{deluxetable*}{ccccccccccccccc}
\tabletypesize{\footnotesize}
\tablecolumns{9}
\setlength{\tabcolsep}{1pt}
\tablewidth{0pt}
\tablecaption{Flux densities\label{tab:flux}}
\tablehead{ 
\colhead{Source} & 
\colhead{$F_{\rm 198 GHz}$} &
\colhead{$\delta F_{\rm 198 GHz}$} &
\colhead{$F_{\rm 220 GHz}$} &
\colhead{$\delta F_{\rm 220 GHz}$} &
\colhead{$F_{\rm 227 GHz}$} &
\colhead{$\delta F_{\rm 227 GHz}$} &
\colhead{$F_{\rm 249 GHz}$} &
\colhead{$\delta F_{\rm 249 GHz}$} &
\colhead{$F_{\rm 336 GHz}$} &
\colhead{$\delta F_{\rm 336 GHz}$} &
\colhead{$F_{\rm 358 GHz}$} &
\colhead{$\delta F_{\rm 358 GHz}$} &
\colhead{$F_{\rm 407.5 GHz}$} &
\colhead{$\delta F_{\rm 407.5 GHz}$} \\
&
(mJy) & (mJy) & (mJy) & (mJy) & (mJy) & (mJy) & (mJy) & (mJy) & (mJy) & (mJy) & (mJy) & (mJy) & (mJy) & (mJy)
}
\startdata 
CIDA-1 & 14.6 & 2.0 &   16.0 & 2.4 & 16.5 & 2.2 & 20.0 & 2.7 & 33.6 & 3.1 & 39.7 & 5.0  & 63.1 & 21.6 \\
FZ Tau & 13.4 & 2.4 &   16.9 & 2.9 & 18.8 & 2.6 & 12.8 & 3.1 & 28.9 & 3.1 & 25.4 & 4.9  & 37.8 & 18.7 \\
V410 X-ray 2 & 10.3 & 2.1 &   11.9 & 2.5 & 12.5 & 2.3 & 18.8 & 2.8 & 24.9 & 3.3 & 33.7 & 5.7  & 69.9 & 25.0 \\
CX Tau & 8.0 & 1.4 &   7.5 & 1.7 & 10.1 & 1.5 & 11.9 & 1.8 & 22.0 & 2.3 & 30.0 & 3.8  & 47.0 & 15.6 \\
V807 Tau & 7.3 & 1.6 &   7.8 & 1.9 & 8.6 & 1.7 & 10.9 & 2.1 & 18.9 & 2.3 & 25.1 & 3.9  & 8.9 & 10.3 \\
FX Tau & 7.3 & 1.2 &   6.4 & 1.4 & 10.6 & 1.3 & 8.7 & 1.5 & 14.6 & 1.6 & 19.3 & 2.7  & 39.9 & 10.2 \\
IT Tau A & 3.7 & 1.2 &   7.4 & 1.3 & 8.0 & 1.4 & 12.1 & 1.6 & 15.9 & 1.7 & 19.0 & 2.6  & 36.2 & 9.6 \\
IT Tau B & 2.8 & 1.1 &   5.5 & 1.2 & 3.4 & 1.3 & 6.1 & 1.5 & 8.2 & 1.6 & 12.9 & 2.5  & 19.6 & 9.2 \\
KPNO 10 & 5.1 & 1.1 &   5.5 & 1.3 & 5.4 & 1.2 & 7.8 & 1.5 & 14.8 & 1.8 & 18.3 & 3.1  & 54.0 & 12.2 \\
04301+2608 & 3.5 & 1.3 &   4.0 & 1.6 & 4.8 & 1.3 & 3.5 & 1.5 & 9.1 & 1.5 & 9.3 & 2.7  & 16.7 & 9.4 \\
\enddata
\tablecomments{Flux densities ($F$) of the observed flux densities and the 1-$\sigma$ thermal noises ($\delta F$), which were derived based on fitting the complex visibilities (Section \ref{sub:flux density}). The measurements were made after applying the absolute flux rescaling factors introduced in Appendix \ref{appdx:flux}.}
\end{deluxetable*}

\bibliography{main}{}
\bibliographystyle{aasjournal}

\appendix

\section{Absolute flux scales}\label{appdx:flux}

We fit power-laws to the flux densities of IC~2087~IR and V892~Tau quoted from \citet{Chung2024ApJS..273...29C}.
We took these power-laws as the absolute flux models, which are more immune to the uncertainties of absolute flux calibrations in individual tracks of the observations.

To derive the absolute flux rescaling factors ($C(\nu)$) for the observations listed in Table \ref{tab:obs_summary}, we compared the flux densities of IC~2087~IR and V892~Tau measured from individual of these tracks of observations (c.f. Section \ref{sub:flux density}) with the power-laws mentioned above.
For each track of observations, $C(\nu)$ is defined as what minimizes
\[
\frac{(C(\nu)F^{\rm IC2087IR}_{\nu} - f^{\rm IC2087IR}_{\nu})^2}{ (\delta F^{\rm IC2087IR}_{\nu})^2 } + \frac{(C(\nu)F^{\rm V892Tau}_{\nu} - f^{\rm V892Tau}_{\nu})^2}{ (\delta F^{\rm V892Tau}_{\nu})^2 },
\]
where $F^{\rm IC2087IR}_{\nu}$ is the flux density of IC~2087~IR measured from that track and at a frequency $\nu$, $\delta F^{\rm IC2087IR}_{\nu}$ is the 1-$\sigma$ uncertainty of that flux density measurement, and $f^{\rm IC2087IR}_{\nu}$ is the flux density inferred from the absolute flux models described above, etc.

\begin{figure}
\hspace{-1.6cm}
  \begin{tabular}{ cc }
    \includegraphics[width=4.7cm]{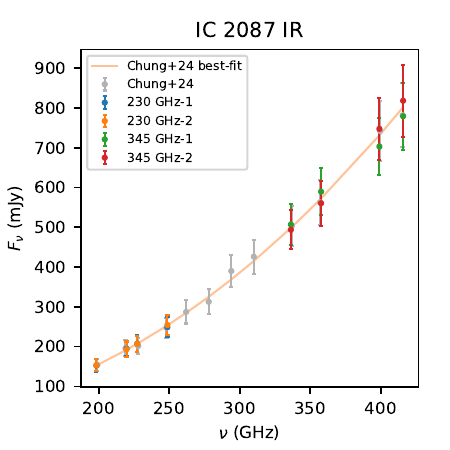}   &
    \includegraphics[width=4.7cm]{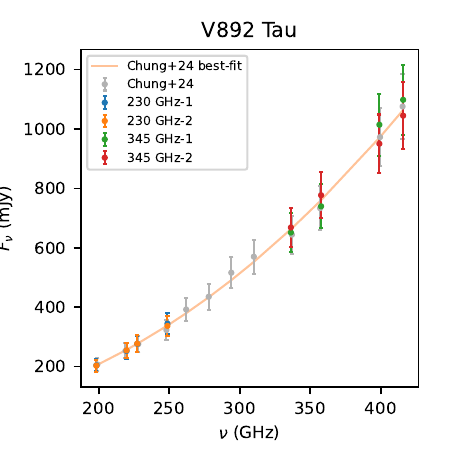}   \\ 
  \end{tabular}
  \caption{Flux densities of the two calibration sources, IC~2087~IR and V892~Tau. Color symbols show the measurements from the new SMA observations introduced in Section \ref{sec:observation}, after rescaling based on the procedure outlined in Section \ref{sub:flux density}; gray symbols and solid lines show the measurements quoted from \citet{Chung2024ApJS..273...29C} and the best-fit power-laws for the quoted measurements. 
  }
  \label{fig:calData}
\end{figure}

We listed the derived values of $C(\nu)$ in Table \ref{tab:rescale}.
The corrections are mostly only a few precent at 198--358 GHz. 
The corrections are $\sim$30\%--50\% at higher frequencies owning to the weather conditions.
Figure \ref{fig:calData} shows the corrected flux densities of IC~2087~IR and V892~Tau.

\begin{deluxetable}{l c c}[h]
\tabletypesize{\scriptsize}
\tablecolumns{3}
\tablewidth{6cm}
\tablecaption{Rescaling factors\label{tab:rescale}}
\tablehead{ 
\colhead{Track ID} & 
\colhead{Central Freq.} &
\colhead{Rescaling factor} \\
& (GHz) & 
}
\startdata 
\multirow{4}{*}{230 GHz-1}   & 198 & 0.999 \\
   & 220 & 1.035 \\
   & 227 & 1.007 \\
   & 249 & 1.039 \\\hline
\multirow{4}{*}{230 GHz-1}   & 198 & 0.996 \\
   & 220 & 1.029 \\
   & 227 & 0.999 \\
   & 249 & 1.032 \\\hline
\multirow{4}{*}{345 GHz-1}   & 336 & 1.109 \\
   & 358 & 1.070 \\
   & 398 & 1.537 \\
   & 418 & 1.471 \\\hline
\multirow{4}{*}{345 GHz-2}   & 336 & 1.037 \\
   & 358 & 1.07 \\
   & 398 & 1.26 \\
   & 418 & 1.31 \\
\enddata
\tablecomments{The rescale factors for flux densities in the individual tracks of observations (Appendix \ref{appdx:flux}). The second column lists the central frequencies at each sideband of each of the two SMA receivers (Section \ref{sub:obs}).}
\vspace{-1cm}
\end{deluxetable}

\section{Posteriors derived from the fittings of spectral indices}\label{appdx:posterior}

\begin{figure*}
    \hspace{-1cm}
    \begin{tabular}{cc}
    \includegraphics[width=8.5cm]{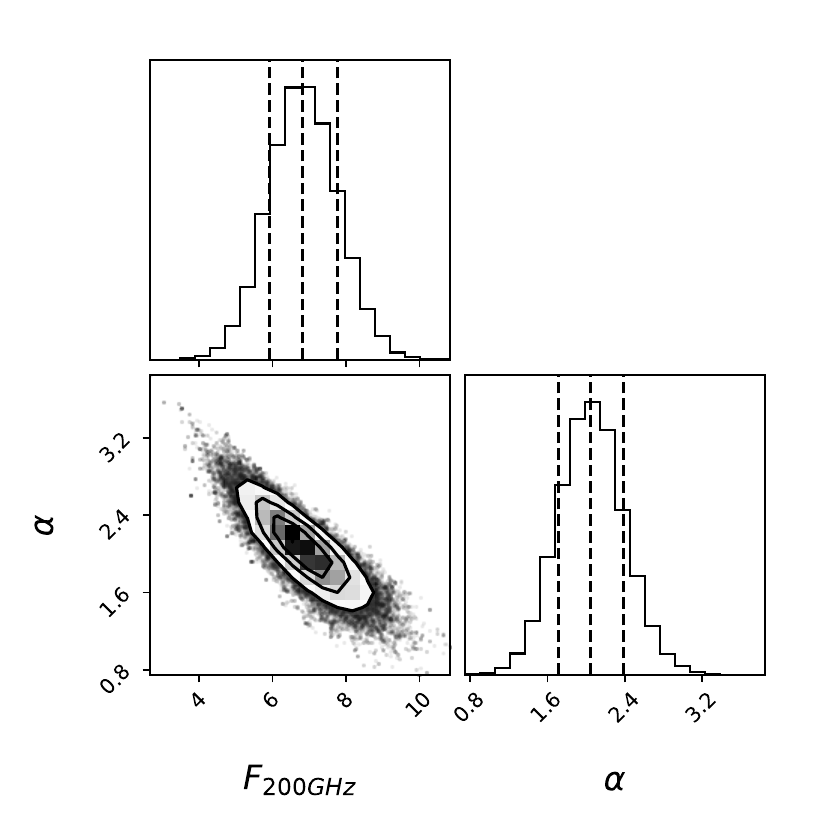} &
    \includegraphics[width=8.5cm]{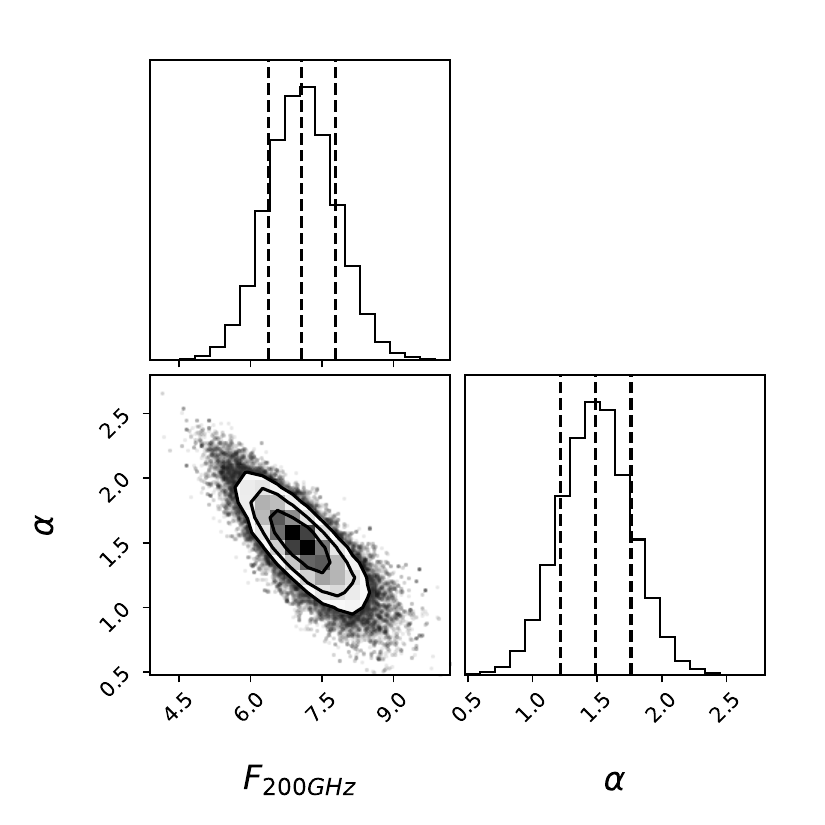} \\
    \end{tabular}
    \caption{The corner plots for V807~Tau (left) and FX~Tau (right), which show the posterior distributions of the flux densities at 200 GHz ($F_{\rm 200GHz}$) and spectral indices ($\alpha$) derived in the power-law fittings (\ref{sub:SED fitting}).
    }
    \label{fig:corner}
\end{figure*}

Figure \ref{fig:corner} provides the examples of the corner plots generated when fitting power-laws to the observed spectra (Section \ref{sub:SED fitting}).
The flux densities of the presented two sources V807~Tau and FX~Tau are close to the median value of our sample (Table \ref{tab:flux}).

\section{Disk radii}\label{appdx:radius}

Except for CIDA~1, the objects detected in our SMA observations are spatially compact. 
In the previous, higher angular resolution Atacama Large Millimeter/submillimeter Array (ALMA) observations, the dust image of CIDA~1 is composed of a ring and an unresolved inner disk (\citealt{Pinilla2018A&A...615A..95P,Kurtovic2021A&A...645A.139K,Pinilla2021A&A...649A.122P,Hashimoto2023AJ....166..186H}).
We quoted the $R_{90\%}$ value of this source from \citet{Kurtovic2021A&A...645A.139K}, which is a good approximation for $R_{95\%}$.

\begin{deluxetable*}{llccc}
\tabletypesize{\footnotesize}
\tablecolumns{5}
\tablewidth{0pt}
\tablecaption{Data for deriving the $R_{\rm 95\%}$ radii\label{tab:radii}}
\tablehead{ 
\colhead{Source} & ALMA project & \colhead{ALMA synthesized beam} & Frequency & \colhead{Deconvolved FWHM}  \\
    & & ($''\times''$; $^{\circ}$) & (GHz) & (mas$\times$mas)
}
\startdata 
FZ Tau & 2021.1.00854.S (PI: Long, F.) & 0.073$\times$0.055; $-$28 & 225.5 & $(81\pm2.9)\times(69\pm3.2)$ \\
CX Tau & 2016.1.00715.S (PI: Facchini, S.) & 0.058$\times$0.032; 13 & 225.5 & $(135\pm5.8)\times(72\pm4.1)$ \\
V807 Tau & 2013.1.00105.S (PI: Akeson, R.) & 0.21$\times$0.15; 25 & 237.5 & $(91\pm14)\times(78\pm23)$ \\
FX Tau & 2013.1.00105.S (PI: Akeson, R.) &  0.21$\times$0.14; 25 &  237.5 & $(158\pm19)\times(100\pm26)$ \\
IT Tau A & 2022.1.01302.S (PI: Mulders, G.) & 0.35$\times$0.24; $-$0.96 & 257.0 & $(187\pm15)\times(130\pm19)$ \\
IT Tau B & 2022.1.01302.S (PI: Mulders, G.) & 0.35$\times$0.24; $-$0.96 & 257.0 & $(139\pm23)\times(126\pm30)$ \\
KPNO 10 &  2016.1.01511.S (PI: Patience, J.) & 0.19$\times$0.11; $-$42 & 338.8 & $(133\pm10)\times(108\pm13)$ \\
\enddata
\end{deluxetable*}

There are archival, $\lesssim$0\farcs3 ($\lesssim$55 au in terms of FWHM) angular resolution ALMA observations on IT~Tau~A, IT~Tau~B, KPNO~10, V807~Tau, FX~Tau, CX~Tau, and FZ~Tau at 225--338 GHz.
For these objects, we downloaded the QA2-processed ALMA images and then performed two-dimensional Gaussian fittings in the image domain. 
We approximated their $R_{\rm 95\%}$ radii by multiplying the deconvolved FWHM in the major axis by 1.04.
The ALMA data we utilized and the deconvolved angular scales of these objects are summarized in Table \ref{tab:radii}.

At the time this paper is written, there are no high angular resolution ALMA observations of the two objects, 04301+2608 and V410~X-ray~2.
Using Equation (2) of \citet{Hendler2020ApJ...895..126H}, which was the measured luminosity-radius relation (see also \citealt{Tripathi2017ApJ...845...44T,Andrews2018ApJ...869L..41A}), we estimated the $R_{\rm 68\%}$ for these two objects based on their millimeter luminosity.
We then used Equation (6) of \citet{Hendler2020ApJ...895..126H} to convert from $R_{\rm 68\%}$ to $R_{\rm 90\%}$, as a good approximations of their $R_{\rm 95\%}$.



\end{document}